\documentclass[pra,10pt,a4paper,twocolumn,nofootinbib, superscriptaddress]{revtex4-1}
\usepackage{graphicx}
\usepackage{amsmath}
\usepackage{amsfonts}
\usepackage{amssymb}
\usepackage{amstext} 
\usepackage{soul}
\usepackage{dsfont}
\usepackage[utf8]{inputenc}
\usepackage{hyperref}
\usepackage{color}

\begin{document}
\title{General solution of the time evolution of two interacting harmonic oscillators}
\date{\today}

\author{David Edward Bruschi}
\email{david.edward.bruschi@posteo.net}
\affiliation{Theoretical Physics, Universit\"at des Saarlandes, 66123 Saarbr\"ucken, Germany}
\altaffiliation{Present address: Institute for Quantum Computing Analytics (PGI-12), Forschungszentrum J\"ulich, 52425 J\"ulich, Germany}
\affiliation{Central European Institute of Technology (CEITEC), Brno University of Technology, 621 00 Brno, Czech Republic}
\author{G. S. Paraoanu}
\affiliation{QTF Centre of Excellence,  Department of Applied Physics, Aalto University School of Science, FI-00076 AALTO, Finland}
\author{Ivette Fuentes A}
\affiliation{School of Mathematical Sciences, University of Nottingham, NG7 2RD, Nottingham, UK}
\altaffiliation{{Present address: School of Physics and Astronomy, University of Southampton, SO17 1BJ Southampton, UK}}
\author{Frank K. Wilhelm}
\affiliation{Theoretical Physics, Universit\"at des Saarlandes, 66123 Saarbr\"ucken, Germany}
\altaffiliation{Present address: Institute for Quantum Computing Analytics (PGI-12), Forschungszentrum J\"ulich, 52425 J\"ulich, Germany}
\author{Andreas W. Schell}
\affiliation{Institut für Festkörperphysik, Leibniz Universität Hannover, 30167 Hannover, Germany}
\affiliation{Physikalisch-Technische Bundesanstalt, 38116 Braunschweig, Germany}
\affiliation{Central European Institute of Technology (CEITEC), Brno University of Technology, 621 00 Brno, Czech Republic}

\begin{abstract}
We study the time evolution of an ideal system composed of two harmonic oscillators coupled through a quadratic Hamiltonian with arbitrary interaction strength. We solve its dynamics analytically by employing tools from symplectic geometry. In particular, we use this result to completely characterize the dynamics of the two oscillators interacting in the ultrastrong coupling regime with additional single-mode squeezing on both oscillators, as well as higher order terms. Furthermore, we compute quantities of interest, such as the average number of excitations and the correlations that are established between the two subsystems due to the evolution. We find that this model predicts a second order phase transition and we compute the critical exponents and the critical value. We also provide an exact decoupling of the time evolution in terms of simple quantum optical operations, which can be used for practical implementations and studies. Finally, we show how our techniques can be extended to include more oscillators and higher order interactions. 
\end{abstract}

\maketitle

\section{Introduction}
Harmonic oscillators are paramount among quantum systems, due to their great importance for the description of bosonic systems, such as the modes of the electromagnetic field or phononic excitations of optomechanical systems \cite{Aspelmeyer:Kippenberg:2014}, many body systems \cite{Dalfovo:Giorgini:1999}, circuit quantum electrodynamics (QED) \cite{Blais:Huang:2004}, Dicke models in the thermodynamical limit \cite{Emary:Brandes:2003,Emary:Brandes:2003:v2,Cuccoli:Giachetti:1995}, and cavity QED \cite{Haroche:Raimond:2010}.
Regardless of the overwhelming number of studies that have used coupled harmonic oscillators for different tasks, there is still a lack of analytical control and understanding of relatively simple systems, such of two coupled harmonic oscillators in the ultrastrong coupling regime. Some progress in this direction has been achieved for oscillators in the ultrastrong coupling regime on resonance \cite{Estes:Keil:1968}, for charged particles in anisotropic potentials \cite{Dippel:Schmelcher:1994}, and by using higher order perturbation theory \cite{Mukhopadhyay:2018}, while approaches to general solutions have been put forward providing results of different degrees of complexity \cite{Bruschi:Lee:2013,Sandoval-Santana:Ibarra-Sierra:2016}. In general, an explicit solution to the full problem remains outstanding.

In this work we close this gap and provide the exact solution of the time evolution of a system of two coupled harmonic oscillators that interact through a time independent quadratic Hamiltonian. Our results are free from approximations and apply to the whole parameter space. Therefore, these results allow us to study the dynamics of the harmonic oscillators on a variety of experimental platforms. We apply these solutions to the important scenario of two oscillators that interact in the ultrastrong coupling regime with the addition of single-mode squeezing on each oscillator. The solutions are given explicitly in terms of the Hamiltonian parameters, and we employ these results to compute a few quantities of physical interest, such as particle creation number and entropy of entanglement between the modes.  We also discuss the occurrence of phase transitions for which we are able to find the critical value and the critical exponents, and we highlight which existing physical systems can benefit from this analytical understanding \cite{Emary:Brandes:2003,Baksic:Ciuti:2014}. This analytical control over the dynamics can provide new insights and motivate the pursuit of new experimental regimes of operation.  We also discuss extensions of this work to tackle diagonalization of Hamiltonians of higher order, i.e., Hamiltonians that contain cubic or higher powers of the quadrature operators \cite{Fuentes:Barberis-Blostein:2007}. These terms appear, for example, in statistical mechanics \cite{Cuccoli:Giachetti:1995} and models of interacting molecules \cite{Zamastil:Cizek:2000,Liverts:Mandelzweig:2006}.
Furthermore, we use mathematical tools from symplectic geometry to show how the time evolution induced by the ultrastrong coupling Hamiltonian, with additional single-mode squeezing, can be decomposed into a sequence of \textit{time independent} simple quantum optical operations, i.e., mode mixing and squeezing, while the dependence on time enters only through a phase rotation throughout the process. This result allows us to interpret the time evolution as a channel, which can be implemented through basic operations either in a simulation or in an experiment. This decomposition can open the door to applying and combining our results to other fields, such as the theory of quantum channels \cite{quantuminfo}. 

This paper is organized as follows. In Section \ref{section:tools} we introduce the necessary tools. In Section \ref{time:evolution:section} we solve the time evolution analytically and we characterize the phase transition. In Section \ref{useful:quantities:section} we apply our results to compute a few relevant quantities, such as the entropy of entanglement. Section \ref{section:higher:order} deals with the introduction of higher order terms. Finally, in Section \ref{circuit:section} we provide a decoupling of the time evolution in terms of simple quantum optical operations.

\section{Tools}\label{section:tools}

\subsection{Hamiltonian}
A generic quadratic Hamiltonian describing two harmonic oscillators with mass $m_k$, momentum $\hat{p}_k$, position $\hat{x}_k$ and frequency $\omega_k$, where we have $k=$a,b, is given by
{\small
\begin{align}\label{basic:hamiltonian:physical}
\hat{H}=\frac{1}{2}\left[\frac{\hat{p}^2_\text{a}}{m_\text{a}}+\omega_\text{a}^2\hat{x}^2_\text{a}+\frac{\hat{p}^2_\text{b}}{m_\text{b}}+\omega_\text{b}^2\hat{x}^2_\text{b}\right]+ V(\hat{x}_\text{a},\hat{x}_\text{b},\hat{p}_\text{a},\hat{p}_\text{b}),
\end{align}
}
where $V(\hat{x}_\text{a},\hat{x}_\text{b},\hat{p}_\text{a},\hat{p}_\text{b})$ is a potential formed by quadratic combinations of the position and momentum operators, and the canonical commutation relations read $[\hat{x}_k,\hat{p}_{k'}]=i\,\hbar\delta_{kk'}$. The Hamiltonian \eqref{basic:hamiltonian:physical} can be written in terms of creation and annihilation operators 
\begin{align}
\hat{a}=&\frac{1}{\sqrt{2}}(\hat{q}_{\text{a}}+i\hat{p}_\text{a}) &
\hat{a}^\dagger=&\frac{1}{\sqrt{2}}(\hat{q}_{\text{a}}-i\hat{p}_{\text{a}})\nonumber\\
\hat{b}=&\frac{1}{\sqrt{2}}(\hat{q}_\text{b}+i\hat{p}_\text{b}) &
\hat{b}^\dagger=&\frac{1}{\sqrt{2}}(\hat{q}_\text{b}-i\hat{p}_\text{b}),
\end{align}
where we have introduced the quadrature operators $\hat{q}_\text{a}:=\sqrt{m_\text{a}\,\omega_\text{a}/\hbar}\,\hat{x}_\text{a}$ and $(\hbar\,m_\text{a}\,\omega_\text{a})^{-1/2}\,\hat{p}_\text{a}\rightarrow\hat{p}_\text{a}$, and analogous for mode $\hat{b}$. In this new basis, the Hamitonian \eqref{basic:hamiltonian:physical} takes the form
\begin{align}\label{second:basic:hamiltonian:physical}
\hat{H}=\hat{H}_0+V(\hat{a},\hat{b},\hat{a}^\dag,\hat{b}^\dag),
\end{align}
where $\hat{H}_0=\hbar\,\omega_\text{a}\,\hat{a}^\dag\,\hat{a}+\hbar\,\omega_d\,\hat{b}^\dag\,\hat{b}$ denotes the free Hamiltonian (we have discarded the zero point energy constant), and now the potential $V$ is obtained as a combination of terms that are quadratic in the new operators.   

The potential can take, for example, the form $V(\hat{a},\hat{b},\hat{a}^\dag,\hat{b}^\dag)=g\,(\hat{a}\,\hat{b}^\dag+\hat{a}^\dag\,\hat{b}+\hat{a}\,\hat{b}+\hat{a}^\dag\,\hat{b}^\dag)$, and in this case we see that the first two terms take excitations of one oscillator into the other one (mode mixing) while the second two terms are associated with parametric amplification (squeezing). The system described by \eqref{second:basic:hamiltonian:physical} with such potential is said to be in the ultrastrong coupling regime  \cite{Kockum:Miranowicz:2019,Diaz:Lamata:2019} if the coupling strength $g$ is comparable with the frequencies $\omega_\text{a}$ and $\omega_\text{d}$.
An extension of this ultrastrong coupling Hamiltonian \eqref{second:basic:hamiltonian:physical} occurs when mode mixing and squeezing have different coupling constants. In addition, if we include also single-mode squeezing of each oscillator, we obtain the Hamiltonian of the form
\begin{align}\label{basic:hamiltonian}
\hat{H}=&\hat{H}_0+\hbar\,g_\text{bs} \left(\hat{a}\,\hat{b}^\dag+\hat{a}^\dag\,\hat{b}\right)+\hbar\,g_\text{sq} \left(\hat{a}^\dag\,\hat{b}^\dag+\hat{a}\,\hat{b}\right)\nonumber\\
&+\hbar\,\lambda_\text{a}\,\left(\hat{a}^{\dag2}+\hat{a}^2\right)+\hbar\,\lambda_\text{b}\,\left(\hat{b}^{\dag2}+\hat{b}^2\right).
\end{align}
Here, $g_\text{bs}$, $g_\text{sq}$, $\lambda_\text{a}$, and $\lambda_\text{b}$  are the coupling constants of the mode mixing interaction, the squeezing interaction, the single-mode squeezing of mode $\hat{a}$, and the single-mode squeezing of mode $\hat{b}$ respectively.

Hamiltonians of this form can be engineered in physical systems such as coupled nanomechanical oscillators \cite{Monteiro:Millen:2013,Aspelmeyer:Kippenberg:2014}, coupled vibrational modes of molecules \cite{delPino:Feist:2015} and coupled microwave resonators in circuit QED \cite{Jin:Rossini:2013,Jin:Rossini:2014}, see also Appendix \ref{superconducting:circuits:appendix}.

\subsection{Time evolution}
The time evolution operator $\hat{U}(t)$ induced by a time-dependent Hamiltonian $\hat{H}(t)$ reads
\begin{align}\label{time:evolution:operator}
\hat{U}=\overset{\leftarrow}{\mathcal{T}}\exp\left[-\frac{i}{\hbar}\int_0^{t}\,dt'\,\hat{H}(t')\right],
\end{align}
where $\overset{\leftarrow}{\mathcal{T}}$ stands for the time-ordering operator.

The solution to the implicit expression \eqref{time:evolution:operator} for our Hamiltonian \eqref{basic:hamiltonian} has already been obtained when $\omega_\text{a}=\omega_\text{d}$, that is, on resonance \cite{Estes:Keil:1968}. Further analysis of this system within higher order perturbation theory has also given partial results \cite{Mukhopadhyay:2018}. In this work we proceed beyond this special case, and provide the full solution to the problem.

\subsection{Linear dynamics}
The Hamiltonian \eqref{basic:hamiltonian} induces \textit{linear dynamics}.\footnote{In this work we use the term \textit{linear} for unitary operators induced by Hamiltonians that are quadratic in the creation and annihilation operators. This terminology is completely independent from the fundamental linearity of quantum mechanics.} Such dynamics are defined as those induced by Hamiltonians that are \textit{quadratic} in the creation and annihilation operators (or equivalently in the quadrature operators), which is the case of this work. Therefore, we can use the symplectic formalism \cite{Adesso:Ragy:2014} to map the usually intractable problem of manipulating unitary operators to the much more tractable problem of multiplying low-dimensional matrices, an approach that has been recently developed in the literature for linear systems \cite{Bruschi:Lee:2013} -- an alternative approach was attempted in \cite{Brown:Matrin-Martinez:2013}, and has also been extended to nonlinear ones \cite{Bruschi:Xuereb:2018}. An extensive review can be found in the literature \cite{Adesso:Ragy:2014}. 

We start by collecting the creation and the annihilation operators in the vector $\hat{\mathbb{X}}$ of operators defined as $\hat{\mathbb{X}}:=(\hat{a},\hat{b},\hat{a}^\dag,\hat{b}^\dag)^{\text{Tp}}$.
Any linear unitary evolution of our two oscillators can be represented by a $4\times4$ \textit{symplectic} matrix $\boldsymbol{S}$ through the defining equation 
\begin{align}\label{time:evolution:operator:symplectic:representation}
\hat{\mathbb{X}}(t)=\hat{U}(t)^\dag\,\hat{\mathbb{X}}\,\hat{U}(t)=\boldsymbol{S}\,\hat{\mathbb{X}}.
\end{align}
The defining property of a symplectic matrix $\boldsymbol{S}$ is that it satisfies $\boldsymbol{S}\,\boldsymbol{\Omega}\,\boldsymbol{S}^\dag=\boldsymbol{S}^\dag\,\boldsymbol{\Omega}\,\boldsymbol{S}=\boldsymbol{\Omega}$, where $\boldsymbol{\Omega}$ is the \textit{symplectic form} \cite{Adesso:Ragy:2014}. Given the choice of ordering of the operators in the vector $\hat{\mathbb{X}}$, we have that
\begin{align}
\boldsymbol{S}
=
\begin{pmatrix}
\boldsymbol{\alpha} & \boldsymbol{\beta}\\
\boldsymbol{\beta}^* & \boldsymbol{\alpha}^*
\end{pmatrix},
\end{align}
and $\boldsymbol{\Omega}=-i\,\textrm{diag}(1,1,-1,-1)$. Notice that the defining property of the symplectic matrix $\boldsymbol{S}$ is equivalent to the well-known Bogoliubov identities, which in matrix form read $\boldsymbol{\alpha}\,\boldsymbol{\alpha}^\dag-\boldsymbol{\beta}\,\boldsymbol{\beta}^\dag=\mathds{1}$ and $\boldsymbol{\alpha}\,\boldsymbol{\beta}^{\text{Tp}}-\boldsymbol{\beta}\,\boldsymbol{\alpha}^{\text{Tp}}=0$.

Finally, any quadratic Hamiltonian $\hat{H}$ can be put in a matrix form $\boldsymbol{H}$ by the following
\begin{align}\label{symplectic:representation:hamiltonian}
\hat{H}=\frac{\hbar}{2}\hat{\mathbb{X}}^\dag\,\boldsymbol{H}\,\hat{\mathbb{X}}, \quad\quad 
\boldsymbol{H}
=
\begin{pmatrix}
\boldsymbol{U} & \boldsymbol{V} \\
\boldsymbol{V}^* & \boldsymbol{U}^*
\end{pmatrix},
\end{align}
where $\boldsymbol{U}$ and $\boldsymbol{V}$ satisfy $\boldsymbol{U}=\boldsymbol{U}^\dag$ and $\boldsymbol{V}=\boldsymbol{V}^T$.

Therefore, the action \eqref{time:evolution:operator:symplectic:representation} of the time evolution operator $\hat{U}(t)$ on the (vector of) creation and annihilation operators implies that it has the \textit{symplectic representation} $\boldsymbol{S}(t)$ of the form
\begin{align}\label{symplectic:representation:time:evolution:operator}
\boldsymbol{S}(t)=\overset{\leftarrow}{\mathcal{T}}\,\exp\left[\boldsymbol{\Omega}\,\int_0^t\,dt'\,\boldsymbol{H}(t')\right],
\end{align}
which is easy to verify explicitly.

We note here that Bogoliubov transformations are symplectic transformations. In fact, the defining properties of symplectic matrices are just another way of stating that the transformations preserve the canonical commutation relations.

\subsection{Covariance matrix formalism}
In this work we will consider Gaussian states of bosonic systems. Gaussian states are prominent across many areas of physics \cite{Adesso:Ragy:2014}. In conjunction with the techniques defined above, they allow for a full description and characterisation of the whole physical system using the covariance matrix formalism. Note that, while the analytical solution of the time evolution applies to systems that are initially in any state, the addition of the covariance matrix formalism can be done only when considering Gauissian states. A full introduction to this topic can be found in the literature \cite{Adesso:Ragy:2014}. 

A Gaussian state $\hat{\rho}_\text{G}$ of $N$ bosonic modes in the covariance matrix formalism is \textit{fully} characterised by the $N$-dimensional vector of first moments $d$ and the $N\times N$ covariance matrix of second moments $\boldsymbol{\sigma}$ defined by
\begin{align}
d_n:=&\langle \hat{X}_n\rangle\nonumber\\
\sigma_{nm}:=&\langle\hat{X}_n\hat{X}_m^\dag+\hat{X}_m^\dag\hat{X}_n\rangle-2\,\langle \hat{X}_n\rangle\langle \hat{X}_m^\dag\rangle.
\end{align}
Here, $\hat{X}_n$ is the $n$-th element of the vector $\hat{\mathbb{X}}$ of operators, $\langle\cdot\rangle$ is the average with respect to the state $\hat{\rho}_\text{G}$.

We have seen that if a unitary transformation $\hat{U}(t)$ acting on an initial Gaussian state is \textit{linear}, then it is represented by a symplectic matrix $\boldsymbol{S}$ through \eqref{symplectic:representation:time:evolution:operator}, and the usual von Neumann relation $\hat{\rho}_\text{G}(t)=\hat{U}(t)\,\hat{\rho}_\text{G}(0)\,\hat{U}^\dag(t)$ takes the form
\begin{align}\label{covariance:matrix:time:evolution}
\boldsymbol{\sigma}(t) = &\boldsymbol{S}(t)\,\boldsymbol{\sigma}(0)\,\boldsymbol{S}^\dag(t).
\end{align}
This equation must be supplemented by the trasformation of the first moments, which reads $d(t)=\boldsymbol{S}\,d(0)$.

Williamson's theorem guarantees that any $2\,N\times2\,N$ matrix, such as the covariance matrix $\boldsymbol{\sigma}$, can be put in diagonal form as
$\boldsymbol{\sigma} = \boldsymbol{s}\,\boldsymbol{\nu}_\oplus\,\boldsymbol{s}^\dag$ by an appropriate symplectic matrix $\boldsymbol{s}$, see \cite{Williamson:1923}. The diagonal matrix $\boldsymbol{\nu}_\oplus$ is called the \textit{Williamson form} of the covariance matrix $\boldsymbol{\sigma}$ and has the expression $\boldsymbol{\nu}_\oplus=\text{diag}(\nu_1,...,\nu_N,\nu_1,...,\nu_N)$, where $\nu_n\geq1$ are called the \textit{symplectic eigenvalues} of $\boldsymbol{\sigma}$ and are found as the absolute value of the spectrum of $i\,\boldsymbol{\Omega}\,\boldsymbol{\sigma}$. The general expression for such eigenvalues is $\nu_n=\coth\bigl(\frac{\hbar\,\omega_n}{2\,k_\text{B}\,T_n}\bigr)$, where $T_n$ is a local temperature of each subsystem. This is equivalent to the statement that Gaussian states are locally (i.e., in terms of single subsystems) equivalent to thermal states (up to local unitary transformations) Clearly, when $T_n=0$ for all $n$ one has $\boldsymbol{\nu}_\oplus\equiv\mathds{1}$, i.e., the state is pure.

Finally, we recall that, in this formalism, tracing over a subsystem is extremely easy. It is sufficient to delete the rows and columns in the covariance matrix corresponding to the subsystems one wishes to trace out.

\section{Time evolution of the system}\label{time:evolution:section}
In this section we present the explicit expression of the time evolution represented by the transformation \eqref{symplectic:representation:time:evolution:operator}. We leave all details of the calculations to Appendix \ref{time:evolution:strong:coupling:regime} for simplicity of presentation.

\subsection{Time evolution of the system: full coupling}
Let us now proceed with our main computation. We first note that the problem of computing the time evolution with the Hamiltonian \eqref{basic:hamiltonian} has been so far addressed only in certain particular cases, where $\omega_\text{a}=\omega_\text{b}$ and $g_\text{bs}=g_\text{sq}$ in \cite{Estes:Keil:1968}, for $g_\text{sq}=0$ (i.e., the rotating wave approximation) and for $\lambda_\textrm{a}=\lambda_\textrm{b}=0$ in \cite{Portes:Rodrigues:2008,Urzua:Ramos-Prieto:2019}. 
General methods have also been put forward using Lie algebra approaches \cite{Bruschi:Lee:2013,Sandoval-Santana:Ibarra-Sierra:2016}, however, exact solutions are typically difficult to obtain in this way.

Our approach starts by recalling that the Hamiltonian \eqref{basic:hamiltonian} can be easily diagonalised by a Bogoliubov (i.e., \textit{symplectic}) transformation, therefore providing (symplectic) eigenvalues $\kappa_\pm$, namely, the eigenvalues of $i\,\boldsymbol{\Omega}\,\boldsymbol{H}$. Work in this direction has already been done for simpler scenarios \cite{Abdalla:1994,Mukhopadhyay:2018}. In the present case, some algebra allows us to compute $\kappa_\pm$, which have full expression that can be found in \eqref{symplectic:frequencies:appendix}. 

We are now in the position of obtaining the explicit form of the symplectic representation 
\begin{align}\label{symplectic:expression:strong:coupling}
\boldsymbol{S}(t)
=
\begin{pmatrix}
\boldsymbol{A}(t) & \boldsymbol{B}(t) \\
\boldsymbol{B}^*(t) & \boldsymbol{A}^*(t) 
\end{pmatrix}
\end{align}
of the time evolution operator induced by the full Hamiltonian \eqref{basic:hamiltonian}. We find that the
$2\times2$ matrices $\boldsymbol{A}(t)$ and $\boldsymbol{B}(t)$ that have the expression
\begin{align}\label{main:transformation:equation:main:text}
\boldsymbol{A}(t)&=\boldsymbol{\alpha}^\text{Tp}\,e^{-i\,\boldsymbol{\kappa}\,t}\,\boldsymbol{\alpha}-\boldsymbol{\beta}^\text{Tp}\,e^{i\,\boldsymbol{\kappa}\,t}\,\boldsymbol{\beta}\nonumber\\
\boldsymbol{B}(t)&=\boldsymbol{\alpha}^\text{Tp}\,e^{-i\,\boldsymbol{\kappa}\,t}\,\boldsymbol{\beta}-\boldsymbol{\beta}^\text{Tp}\,e^{i\,\boldsymbol{\kappa}\,t}\,\boldsymbol{\alpha},
\end{align}
where $\boldsymbol{A}^{\text{Tp}}(t)=\boldsymbol{A}(t)$, $\boldsymbol{B}^\dag(t)=-\boldsymbol{B}(t)$, and $\boldsymbol{\kappa}=\textrm{diag}(\kappa_+,\kappa_-)$.

Detailed computations on how to obtain the matrices $\boldsymbol{\alpha}$ and $\boldsymbol{\beta}$ explicitly in terms of the hamiltonian parameters are left to Appendix~\ref{time:evolution:strong:coupling:regime}, but we briefly sketch the procedure here. We start with the form \eqref{symplectic:representation:hamiltonian} of the Hamiltonian, where the matrices $\boldsymbol{U}$ and $\boldsymbol{V}$ read
\begin{align}
\boldsymbol{U}
=
\begin{pmatrix}
\omega_\text{a} & g_\text{bs} \\
g_\text{bs} & \omega_\text{b} 
\end{pmatrix},
\,\,\,
\boldsymbol{V}
=
\begin{pmatrix}
\lambda_\textrm{a} & g_\text{sq} \\
g_\text{sq} & \lambda_\textrm{b}
\end{pmatrix}.
\end{align}
We then recall that a $2N\times2N$ symmetric or Hermitian matrix can be put in diagonal form by means of a symplectic matrix $\boldsymbol{s}$ according to Williamson's theorem. Simple algebra allows us to find
\begin{align}\label{main:constraints:equation:main:text}
\boldsymbol{U}&=\boldsymbol{\alpha}^\text{Tp}\,\boldsymbol{\kappa}\,\boldsymbol{\alpha}+\boldsymbol{\beta}^\text{Tp}\,\boldsymbol{\kappa}\,\boldsymbol{\beta}\nonumber\\
\boldsymbol{V}&=\boldsymbol{\alpha}^\text{Tp}\,\boldsymbol{\kappa}\,\boldsymbol{\beta}+\boldsymbol{\beta}^\text{Tp}\,\boldsymbol{\kappa}\,\boldsymbol{\alpha},
\end{align}
which can in principle be inverted to obtain explicitly the coefficients $\alpha_{nm}$ and $\beta_{nm}$.
The expressions that can be found in this way correspond to those found in the literature for $\omega_\text{a}=\omega_\text{b}$, see \cite{Estes:Keil:1968}.  

Given the solution above, the defining equation \eqref{time:evolution:operator:symplectic:representation} implies that the creation and annihilation operators evolve as
\begin{align}\label{main:result:evolution}
\begin{pmatrix}
\hat{a}(t) \\
\hat{b}(t) \\
\end{pmatrix}
=
\boldsymbol{A}(t)
\begin{pmatrix}
\hat{a} \\
\hat{b} 
\end{pmatrix}
+
\boldsymbol{B}(t)
\begin{pmatrix}
\hat{a}^\dag \\
\hat{b}^\dag
\end{pmatrix}.
\end{align}
This is our main result. An explicit solution can be given once the constraints \eqref{main:constraints:equation:main:text} have been explicitly solved, which might be difficult depending on the exact form of $\boldsymbol{U}$ and $\boldsymbol{V}$.

\subsection{Time evolution of the system: ultrastrong coupling}
In the ultrastrong coupling regime \cite{Kockum:Miranowicz:2019,Diaz:Lamata:2019} we typically have $g_\text{bs}=g_\text{sq}=g$, and therefore the full expression \eqref{symplectic:frequencies:appendix} reduces to
\begin{align}\label{symplectic:frequencies}
\kappa_\pm^2=&\frac{1}{2}\left[\omega_\textrm{a}^2+\omega_\textrm{b}^2-\lambda_\textrm{a}^2-\lambda_\textrm{b}^2\pm\left(\left(\omega_\textrm{a}^2-\omega_\textrm{b}^2+\lambda_\textrm{b}^2-\lambda_\textrm{a}^2\right)^2\right.\right.\nonumber\\
&\left.\left.+16\,\left(\omega_\textrm{a}\,\omega_\textrm{b}+\lambda_\textrm{a}\,\lambda_\textrm{b}\right)\,g^2\right)^{1/2}\right]
\end{align}
which, for $\lambda_\textrm{a}=\lambda_\textrm{b}=0$ reads
\begin{align}\label{normal:mode:frequencies:strong:coupling:usual}
\kappa_\pm^2=&\frac{1}{2} \left[(\omega_\text{a}^2+\omega_\text{b}^2)\pm\sqrt{(\omega_\text{a}^2-\omega_\text{b}^2)^2+16 \omega_\text{a} \omega_\text{b}\,g^2}\right]
\end{align}
for $\omega_\text{a}\geq\omega_\text{b}$ and $\omega_\text{a}<\omega_\text{b}$ respectively.

In this case, the Bogoliubov coefficients $\alpha_{nm}$ and $\beta_{nm}$ found in \eqref{main:constraints:equation:main:text} are computed explicitly in Appendix~\ref{time:evolution:strong:coupling:regime:real}. The final expressions reads
\begin{align}\label{alpha:beta:expressions}
 \alpha_{11}&=\frac{\kappa_++\omega_\text{a}-\lambda_\textrm{a}}{2 \sqrt{\kappa_+(\omega_\text{a}-\lambda_\textrm{a})}}\textrm{c}_\theta & \beta_{11}&=-\frac{\omega_\text{a}-\kappa_+-\lambda_\textrm{a}}{2 \sqrt{\kappa_+(\omega_\text{a}-\lambda_\textrm{a})}}\textrm{c}_\theta \nonumber\\
 \alpha_{12}&=\frac{\kappa_++\omega_\text{b}-\lambda_\textrm{b}}{2 \sqrt{\kappa_+(\omega_\text{b}-\lambda_\textrm{b})}}\textrm{s}_\theta & \beta_{12}&=-\frac{\omega_\text{b}-\kappa_+-\lambda_\textrm{b}}{2 \sqrt{\kappa_+(\omega_\text{b}-\lambda_\textrm{b})}}\textrm{s}_\theta \nonumber\\
 \alpha_{21}&=\frac{\kappa_-+\omega_\text{a}-\lambda_\textrm{a}}{2 \sqrt{\kappa_-(\omega_\text{a}-\lambda_\textrm{a})}}\textrm{s}_\theta & \beta_{21}&=-\frac{\omega_\text{a}-\kappa_--\lambda_\textrm{a}}{2\sqrt{\kappa_-(\omega_\text{a}-\lambda_\textrm{a})}}\textrm{s}_\theta \nonumber\\
 \alpha_{22}&=-\frac{\kappa_-+\omega_\text{b}-\lambda_\textrm{b}}{2\sqrt{\kappa_-(\omega_\text{b}-\lambda_\textrm{b})}}\textrm{c}_\theta & \beta_{22}&=-\frac{\omega_\text{b}-\kappa_--\lambda_\textrm{b}}{2\sqrt{\kappa_-(\omega_\text{b}-\lambda_\textrm{b})}}\textrm{c}_\theta,
\end{align}
where we have introduced $\textrm{c}_\theta:=\cos\theta$ and $\textrm{s}_\theta:=\sin\theta$ for simplicity of presentation, while the angle $0\leq2\theta\leq\pi$  is defined through the trigonometric relation 
\begin{align}\label{theta:definition}
\tan(2\,\theta):=&\frac{\sqrt{(\Gamma-(\lambda_\textrm{b}^2-\lambda_\textrm{a}^2))^2-(\omega_\textrm{a}^2-\omega_\textrm{b}^2)^2}}{(\omega_\text{a}^2-\omega_\textrm{b}^2)}
\end{align}
and the quantity
$\Gamma^2:=\left(\omega_\textrm{a}^2-\omega_\textrm{b}^2+\lambda_\textrm{b}^2-\lambda_\textrm{a}^2\right)^2+16\,\left(\omega_\textrm{a}\,\omega_\textrm{b}+\lambda_\textrm{a}\,\lambda_\textrm{b}\right)\,g^2$. Notice that when $\omega_\textrm{a}=\omega_\textrm{b}$ one has to obtain the equivalent formulas to the ones above starting the calculations by imposing $\omega_\textrm{a}=\omega_\textrm{b}$. It is not possible to set $\omega_\textrm{a}=\omega_\textrm{b}$ at the end. The results for this scenario have already been obtained in previous work \cite{Estes:Keil:1968}. 

\subsection{Time evolution of the system: phase transition at the critical coupling}
We are now in the position to make a few considerations on our main results. 

The symplectic eigenvalues \eqref{symplectic:frequencies} of the full Hamiltonian must be real. This implies that the coupling strength $g$ is limited by the critical coupling $g_\text{cr}$ through the equation $|g|\leq g_\text{cr}$, where $g_\text{cr}$ reads
\begin{align}\label{critical:coupling}
g_\text{cr}:=\frac{1}{2}\sqrt{\frac{|(\omega_\text{a}^2-\lambda_\text{a}^2)\,(\omega_\text{b}^2-\lambda_\text{b}^2)|}{\omega_\text{a}\,\omega_\text{b}+\lambda_\text{a}\,\lambda_\text{b}}}
\end{align}
As expected, the expression for the critical coupling \eqref{critical:coupling} matches that for coupled harmonic oscillators in the ultrastrong regime without squeezing, i.e., $g_\text{cr}=\sqrt{\omega_\text{a}\,\omega_\text{b}}/2$ see \cite{Emary:Brandes:2003}.
Therefore, when $\omega_\text{a}=\lambda_\text{a}$, $\omega_\text{b}=\lambda_\text{b}$, or both, there can be no coupling between the oscillators, that is, $g_\text{cr}=0$ and the constraint $|g|\leq g_\text{cr}$ implies $|g|=0$. Note that the existence of such critical value is a consequence of Williamson's theorem. In fact, the theorem requires that the $2N\times2N$ symmetric or Hermitian matrix that one wishes to put in diagonal form be positive definite \cite{Williamson:1923}. Clearly, when one of the symplectic eigenvalues becomes zero, it implies that the original matrix is not positive definite, and therefore it cannot be put in diagonal form through a symplectic transformation. Physically, this means that, once we reach the critical coupling, the system cannot be recast as a collection of (two in our case) uncoupled harmonic oscillators.

We can further compute what happens close to the critical coupling $g_\text{cr}$. We set $g=g_\text{cr}(1-\epsilon)$, where $\epsilon$ can be positive or negative and $|\epsilon|\ll1$, and we obtain the perturbative expansion of the symplectic eigenvalues \eqref{symplectic:frequencies} to first order $\epsilon$. We have, focussing on $\omega_\textrm{a}\geq\omega_\textrm{b}$,
\begin{align}\label{symplectic:frequencies:perturbative}
\kappa_+^2=&\omega_\textrm{a}^2+\omega_\textrm{b}^2-\lambda_\textrm{a}^2-\lambda_\textrm{b}^2-2\,\frac{(\omega_\text{a}^2-\lambda_\text{a}^2)\,(\omega_\text{b}^2-\lambda_\text{b}^2)}{\omega_\textrm{a}^2+\omega_\textrm{b}^2-\lambda_\textrm{a}^2-\lambda_\textrm{b}^2}\,\epsilon\nonumber\\
\kappa_-=&\frac{1}{\sqrt{g_\text{cr}}}\,\sqrt{2\,\frac{|(\omega_\text{a}^2-\lambda_\text{a}^2)\,(\omega_\text{b}^2-\lambda_\text{b}^2)|}{\omega_\textrm{a}^2+\omega_\textrm{b}^2-\lambda_\textrm{a}^2-\lambda_\textrm{b}^2}}\,|g-g_\text{cr}|^{1/2},
\end{align}
where we have made explicit the dependence of $\kappa_-$ on $|g-g_\text{cr}|$.
This implies that among the Bogoliubov coefficients \eqref{alpha:beta:expressions}, those that are proportional to $1/\sqrt{\kappa_-}$ will be proportional to $|\epsilon|^{-1/4}$. Therefore, they grow unboundedly. The  behavior of $\kappa_-$ for $|g-g_\text{cr}|/g_\text{cr}\ll1$ reveals this to be a second order phase transition \cite{Emary:Brandes:2003}. Such transitions are well known in other areas of physics. For example, they occur in systems that are covered under the umbrella of Dicke models, where the total spin of an ensemble of multiple two-level systems is coupled to a common mode of light \cite{Emary:Brandes:2003}, or in superconducting circuits with artificial atoms \cite{Cuccoli:Giachetti:1995}.  
In the limit of large numbers $N$ of atoms (i.e., two-level systems) Dicke models can be mapped using a Holstein-Primakoff transformation into the Hamiltonian \eqref{basic:hamiltonian}, with $\lambda_\textrm{a}=\lambda_\textrm{b}=0$. This makes possible the identification of the phase transition between the normal phase and the superradiant phase by the condition that both eigenvalues are real (when one of them becomes imaginary we have a phase transition). 
\\
A standard way to classify the phase transition is to note that we can write the dependence of $\kappa_-$ on $|g-g_\text{cr}|$ around the critical value $g_\text{cr}$ as $\kappa_-\propto|g-g_\text{cr}|^{z\,\nu}$. In our case we have $\nu=-1/4$, and $z=2$ is known as the dynamical critical exponent \cite{Emary:Brandes:2003,Emary:Brandes:2003:v2}. 

Finally, note that our Hamiltonian provides a richer scenario compared to those considered for $\lambda_\text{a}=\lambda_\text{b}=g_\text{sq}=0$ due to the fact that the parameter space available is larger \cite{Cuccoli:Giachetti:1995}. Concretely, this is a consequence of the fact that the critical value \eqref{critical:coupling} is not determined only by the one parameter, such as the product of the two frequencies when $\lambda_\text{a}=\lambda_\text{b}=g_\text{sq}=0$, but by the combination of four coupling constants that are all independent. Therefore, the phase transition can occur in a very wide landscape of competing free parameters, potentially leading to different physical regimes \cite{Cuccoli:Giachetti:1995}.

\subsection{Time evolution of the system: multimode extension}
We note here that our results are not limited to two modes only. For example, its possible to use our techniques to solve the Hamiltonian of three harmonic oscillators with interaction Hamiltonian $\hat{H}_\text{I}$ of the form
{\small
\begin{align}
\hat{H}_\text{I}= \lambda_{ad}(\hat{q}_\text{a}-\hat{q}_\text{b})^2+\lambda_{dc}(\hat{q}_\text{b}-\hat{q}_\text{c})^2+\lambda_{ca}(\hat{q}_\text{c}-\hat{q}_\text{a})^2.
\end{align}
}
In fact, the expressions \eqref{symplectic:expression:strong:coupling}, \eqref{main:transformation:equation:main:text} and \eqref{main:constraints:equation:main:text}, as well as the main result \eqref{main:result:evolution} apply to systems of arbitrary number $N$ of modes. The only modifications that will occur are the following: i) all matrices have either dimension $2N\times2N$ or $N\times N$, instead of $4\times4$ and $2\times2$ as in the present case; ii) the matrix $\boldsymbol{\kappa}$ that collects the frequencies of the normal modes will read $\boldsymbol{\kappa}:=\text{diag}(\kappa_1,...,\kappa_N)$; iii) the constraint equations necessary for this case are the straightforward extension of those found in \eqref{main:constraints:equation:main:text}.

Notice that, in general, finding the symplectic matrix $\boldsymbol{s}$ (or, equivalently, the Bogoliubov matrices $\boldsymbol{\alpha}$ and $\boldsymbol{\beta}$) that diagonalizes the Hamiltonian matrix $\boldsymbol{H}$ becomes increasingly hard for $N>2$, see Appendix \ref{time:evolution:strong:coupling:regime}. Nevertheless, our procedure can be repeated step by step whenever the case presents itself. 

\section{Excitations and entanglement in the ultrastrong coupling regime}\label{useful:quantities:section}
Our results provide the time evolution of the creation and annihilation operators which can in turn be used to compute the expectation value of any quantity of interest. 
Here we restrict ourselves to the class of Gaussian states, which allow us to showcase the impact of our result through explicit and analytical expressions. Note that our results can be used for any state, including Fock states, but we leave it to future work to extend this work in the direction of arbitrary states.

The covariance matrix of a Gaussian state evolves through equation \eqref{covariance:matrix:time:evolution}, while the vector of first moments $d:=\langle\hat{\mathbb{X}}\rangle$ evolves as $d(t)=\boldsymbol{S}(t)\,d(0)$. At any time $t$, the covariance matrix $\boldsymbol{\sigma}(t)$ of the system has the form
\begin{align}
\boldsymbol{\sigma}
=
\begin{pmatrix}
\boldsymbol{\mathcal{U}}(t) & \boldsymbol{\mathcal{V}}(t) \\
\boldsymbol{\mathcal{V}}^*(t) & \boldsymbol{\mathcal{U}}^*(t)
\end{pmatrix},
\end{align}
where the $2\times2$ matrices $\boldsymbol{\mathcal{U}}(t)$ and $\boldsymbol{\mathcal{V}}(t)$ read
\begin{align}\label{main:final:state:expression}
\boldsymbol{\mathcal{U}}(t)=&\boldsymbol{A}\,\boldsymbol{\mathcal{U}}(0)\,\boldsymbol{A}^*-\boldsymbol{B}\,\boldsymbol{\mathcal{U}}^*(0)\,\boldsymbol{B}^*-\boldsymbol{A}\,\boldsymbol{\mathcal{V}}(0)\,\boldsymbol{B}\nonumber\\
&+\boldsymbol{B}\,\boldsymbol{\mathcal{V}}^*(0)\,\boldsymbol{A}^*\nonumber\\
\boldsymbol{\mathcal{V}}(t)=&\boldsymbol{A}\,\boldsymbol{\mathcal{V}}(0)\,\boldsymbol{A}-\boldsymbol{B}\,\boldsymbol{\mathcal{V}}^*(0)\,\boldsymbol{B}^*-\boldsymbol{A}\,\boldsymbol{\mathcal{U}}(0)\,\boldsymbol{B}^*\nonumber\\
&+\boldsymbol{B}\,\boldsymbol{\mathcal{U}}^*(0)\,\boldsymbol{A}.
\end{align}
The total number of excitations $N(t)$ can be obtained through the expression $N(t)=\frac{1}{4}\text{Tr}(\boldsymbol{\sigma}(t))-1$, which reads
\begin{align}\label{number:of:excitations}
N(t)=&N(0)-\text{Tr}(\boldsymbol{B}^2\,\Re\boldsymbol{\mathcal{U}}(0))-\text{Tr}\Re(\boldsymbol{B}\,\boldsymbol{A}\,\Re\boldsymbol{\mathcal{V}}(0)).
\end{align}

\subsection{Initial vacuum state}
To illustrate the results we can start with the initial vacuum state $|0\rangle$ of the system. The covariance matrix $\boldsymbol{\sigma}(0)$ of this system is just $\boldsymbol{\sigma}(0)=\mathds{1}_4$.

Using the expression \eqref{number:of:excitations} we have $N(t)=|B_{11}(t)|^2+|B_{22}(t)|^2+2\,|B_{12}(t)|^2$. Therefore, we have
\begin{align}
N(t)=&2\,\text{Tr}\left(\boldsymbol{\beta}\,\boldsymbol{\beta}^\text{Tp}\right)+2\,\text{Tr}\left(\boldsymbol{\beta}\,\boldsymbol{\beta}^\text{Tp}\,e^{i\,\boldsymbol{\kappa}\,t}\,\boldsymbol{\beta}\,\boldsymbol{\beta}^\text{Tp}\,e^{-i\,\boldsymbol{\kappa}\,t}\right)\nonumber\\
&-2\,\text{Tr}\Re\left(\boldsymbol{\beta}\,\boldsymbol{\alpha}^\text{Tp}\,e^{i\,\boldsymbol{\kappa}\,t}\,\boldsymbol{\beta}\,\boldsymbol{\alpha}^\text{Tp}\,e^{i\,\boldsymbol{\kappa}\,t}\right).
\end{align}
The creation of particles from the vacuum depends directly on $\boldsymbol{\beta}$, vanishes for $\boldsymbol{\beta}=0$ and is a signature of squeezing processes. Squeezing would not occur if the original Hamiltonian was used within the rotating wave approximation alone, and without single mode squeezing.

Notice that, in general, the number of excitations will be strictly larger than zero, and can approach zero when $\kappa_+=\kappa_-+2\,n\,\pi$, for some appropriate $n\in\mathbb{Z}$.
Furthermore, the number of excitations oscillates with time, which we expect since the time evolution is unitary.

We can also compute some quantities that measure the amount of correlations, or entanglement, generated between the two modes.
The simplest one is the entropy of entanglement $S_{\text{VN}}$, which is just the von Neumann entropy of one of the two reduced systems (note that, since the state is globally pure, the entropy of entanglement of both subsystems has the same value). In the language of the covariance matrix formalism, we have $S_{\text{VN}}:=f_+(\nu)-f_-(\nu)$, where $\nu$ is the symplectic eigenvalue of the reduced state of mode $\hat{a}$, and $f_\pm(\nu):=\frac{\nu\pm1}{2}\ln\frac{\nu\pm1}{2}$.

The reduced state $\boldsymbol{\sigma}_a(t)$ of mode $\hat{a}$ has the expression
{\small
\begin{align}
\boldsymbol{\sigma}_a
=\mathds{1}+
2\begin{pmatrix}
|B_{11}|^2+|B_{12}|^2 & A_{11}B_{11}+A_{12}B_{12} \\
A_{11}^*B_{11}^*+A_{12}^*B_{12}^* & |B_{11}|^2+|B_{12}|^2
\end{pmatrix},
\end{align}
}
and therefore we find
{\small
\begin{align}
\nu=&\sqrt{(1+2|B_{11}|^2+2|B_{12}|^2)^2-4|A_{11}B_{11}+A_{12}B_{12}|^2}.
\end{align}
}
The coefficients $A_{nm}$ and $B_{nm}$ are defined in \eqref{coefficients:total}.

We have entropy of entanglement when $S_{\text{VN}}(\nu(t))>0$, which occurs when $\nu(t)>1$. In fact, the reduced state of mode $\hat{a}$ is initially pure and therefore $\nu(0)=1$. However, due to the entangling nature of the ultastrong coupling Hamiltonian, we have that $\nu(t)>1$ and therefore some quantum correlations are established between the two systems. Note that, since for continuous variables there is no such thing as a `maximally entangled state', the fact that $\nu(t)>0$ guarantees the presence of correlations between the two subsystems, but the actual numerical value of $S_{\text{VN}}(\nu(t))$ does not provide an intuitive understanding on `how much' correlations are present, since no natural scale is available.\footnote{This is in contrast with, say, qubit systems, where maximal entanglement exists and can be found in states such as Bell states. In this case, the entanglement is normalized to take values between $0$ (no entanglement) to $1$ (maximal entanglement).} 

Finally, entanglement can also be computed using the separability criterion of the partial transpose -- i.e., the PPT criterion \cite{Peres:1996,Adesso:Illuminati:2007} -- which for Gaussian states of two modes can be cast as the following procedure.

We take the full state $\boldsymbol{\sigma}(t)$ ate time $t$ and compute the spectrum of $i\,\boldsymbol{\Omega}\,\boldsymbol{P}\,\boldsymbol{\sigma}(t)\,\boldsymbol{P}$ (note that $\tilde{\boldsymbol{\sigma}}(t)=\boldsymbol{P}\,\boldsymbol{\sigma}(t)\,\boldsymbol{P}$ is the partial transpose of the state $\boldsymbol{\sigma}(t)$ in mode $\hat{b}$, in this basis), where the matrix $\boldsymbol{P}$ reads
\begin{align}
\boldsymbol{P}:=
\begin{pmatrix}
1 & 0 & 0 & 0 \\
0 & 0 & 0 & 1\\
0 & 0 & 1 & 0\\
0 & 1 & 0 & 0
\end{pmatrix}.
\end{align}
The spectrum has eigenvalues $+\tilde{\nu}_\pm,-\tilde{\nu}_\pm$, where $0<\tilde{\nu}_-<\tilde{\nu}_+$. These are also called the symplectic eigenvalues of the partial transpose. Recall that the symplectic eigenvalues $\nu_\pm$ of the state are always greater or equal to one. However, the smallest symplectic eigenvalue $\tilde{\nu}_-$ of the partial transpose can be smaller than one. If that is the case, the PPT criterion \textit{for a bipartite system} guarantees that there is entanglement in the state. Furthermore, \textit{all} measures of entanglement are monotonic functions of $\tilde{\nu}_-$, see \cite{Adesso:Ragy:2014}.

The procedure outlined here allows us to compute the  smallest symplectic eigenvalue $\tilde{\nu}_-$ of the partial transpose and therefore detect the presence of entanglement. The explicit result in terms of the parameters of the problem is too cumbersome to be presented here. However, our results allow for immediate numerical analysis of the amount of entanglement present as a function of the parameters of the problem and of time. We leave it to specialized work to perform analysis of this type.
Finally, we note that entanglement for coupled bosonic systems modelled by Hamiltonians such as \eqref{basic:hamiltonian} has been already numerically studied in different works, such as \cite{Dragan:Fuentes:2011}.

It is possible to apply our results to other initial Gaussian states of interest. Given the expression \eqref{main:final:state:expression} for states at some time $t$, it is possible to analyse thermal states, for which $\boldsymbol{\mathcal{U}}(0)=\text{diag}(\nu_+,\nu_-)$ and $\boldsymbol{\mathcal{V}}(0)=0$, and two mode squeezed states, for which $\boldsymbol{\mathcal{U}}(0)=\cosh r\,\mathds{1}_2$ and  $\boldsymbol{\mathcal{V}}(0)=\sinh r\,\boldsymbol{\sigma}_x$.
We choose to leave these cases for more specialized work.

\section{Extension to higher order interactions}\label{section:higher:order}
The techniques developed in this paper can be extended to solve Hamiltonians that include higher order interactions. By this we mean interactions that are at least cubic in the product of the annihilation and creation operators. We will first provide an example that has already been studied in the literature, and then introduce a more general treatment. 

Consider a quadratic two-mode Hamiltonian which only includes a mode-mixing term
\begin{align}\label{operator:BEC1:Hamiltonian}
\hat{H}_{bs}=&\hbar\omega(\,\hat{a}^\dag\,\hat{a}-\hat{b}^\dag\,\hat{b})+\hbar g\,\left(\hat{a}^\dag\,\hat{b}+\hat{b}\,\hat{a}^\dag\right).
\end{align}
This Hamiltonian is obtained by setting $\omega_b=-\omega_a=-\omega$, $g_{bs}=g$ and $g_{\textrm{sq}}=\lambda_\textrm{a}=\lambda_\textrm{b}=0$ and $g_{bs}=0$ in \eqref{basic:hamiltonian}. 
The matrix form \eqref{symplectic:representation:hamiltonian} of the Hamiltonian is
\begin{align}
\boldsymbol{U}
=
\begin{pmatrix}
\omega & g \\
g &-\omega
\end{pmatrix}
\end{align}
and $\boldsymbol{V}=0$, and in the symplectic matrix formalisim  $\hat{H}_{\textrm{bs}}=\frac{\hbar}{2}\,\hat{\mathbb{X}}^\dag_{\theta}\tilde{\boldsymbol{\kappa}}\,\hat{\mathbb{X}}_{\theta}$ where $\hat{\mathbb{X}}_{\theta}=\boldsymbol{s}_{\theta}\hat{\mathbb{X}}$, and $\boldsymbol{s}_{\theta}$ is the symplectic matrix that diagonalizes it. Therefore, the Hamiltonian matrix $\boldsymbol{H}_{\textrm{bs}}=\frac{\hbar}{2}\boldsymbol{s}_{\theta}^\dag \,\tilde{\boldsymbol{\kappa}}\,\boldsymbol{s}_{\theta}$ is diagonalized by the symplectic matrix $\boldsymbol{s}_{\theta}=\boldsymbol{\alpha}\oplus\boldsymbol{\alpha}^{\ast}$, which is given by $\alpha_{nm}$ Bogoliubov coefficients $\alpha_{11}= \alpha_{22}=\cos\theta$, $\alpha_{12}=-\alpha_{21}=\sin\theta$, and $\tan(2\,\theta)=-g/\omega$. All the other Bogoliubov coefficients are zero. The eigenfrequencies $\kappa_\pm$ define the diagonal matrix  $\tilde{\boldsymbol{\kappa}}:=\boldsymbol{\kappa}\oplus\boldsymbol{\kappa}$, where $\boldsymbol{\kappa}:=\text{diag}(\kappa_+,\kappa_-)$ and $\kappa_\pm=\pm\kappa=\pm\sqrt{\omega^2+g^2}$.

Therefore, the diagonal ``free'' Hamiltonian $\hat{H}_0$ is given by $\hat{H}_0=\frac{\hbar}{2}\hat{\mathbb{X}}^\dag\tilde{\boldsymbol{\kappa}}\hat{\mathbb{X}}=\hbar\sqrt{(\omega^2+g^2})(\hat{a}^\dag\hat{a}-\hat{b}^\dag\hat{b})$. 
Now we proceed to include bi-quadratic interactions
\begin{align}\label{operator:BEC2:Hamiltonian}
\hat{H}_{2}=&C_{aaaa}\,\hat{a}^{\dag2}\,\hat{a}^2+C_{bbbb}\,\hat{b}^{\dag2}\,\hat{b}^2+C_{abab}\,\hat{a}^{\dag}\,\hat{b}^\dag\,\hat{a}\,\hat{b}\nonumber\\&+C_{abbb}\,\hat{a}^{\dag}\,\hat{b}^\dag\,\hat{b}\,\hat{b}+C_{abaa}\,\hat{a}^{\dag}\,\hat{b}^\dag\,\hat{a}\,\hat{a}\nonumber\\&+C_{bbab}\,\hat{b}^{\dag}\,\hat{b}^\dag\,\hat{a}\,\hat{b}+C_{aaab}\,\hat{a}^{\dag}\,\hat{a}^\dag\,\hat{a}\,\hat{b}\nonumber\\&+C_{aabb}\,\hat{a}^{\dag}\,\hat{a}^\dag\,\hat{b}\,\hat{b}+C_{bbaa}\,\hat{b}^{\dag}\,\hat{b}^\dag\,\hat{a}\,\hat{a}+\hat{H}_\textrm{cte}.
\end{align}
The total Hamiltonian $\hat{H}$ that includes both quadratic $\hat{H}_{\textrm{bs}}$ and biquadratic $\hat{H}_2$ terms is given by  $\hat{H}=\hat{H}_{bs}+\hat{H}_2$. Analytical solutions to the dynamics of such full Hamiltonian are found by writing the bi-quadratic Hamiltonian \eqref{operator:BEC2:Hamiltonian} as
\begin{align}\label{operator:BEC2:covariance}
\hat{H}_2=\Lambda_2\frac{\hbar}{2}\left(\hat{\mathbb{X}}^\dag_{\theta} \,\tilde{\boldsymbol{\kappa}}\,\hat{\mathbb{X}}_{\theta}\right)^2\,
\end{align}
and solving for the $C$-coefficients in term of the Bogoliubov coefficients. This yields  $C_{aaaa}=C_{bbbb}=\Lambda_2\,\alpha_{11}\alpha_{22}=\Lambda_2\,\cos^2\theta$, $C_{abab}=-\Lambda_2\,(\alpha_{11}\alpha_{22}+\alpha_{12}\alpha_{21})=-2\,\Lambda_2\,\cos(2\,\theta)$, $C_{aabb}=C_{bbaa}=-\Lambda_2\alpha_{12}\alpha_{21}=\Lambda_2\sin^2\theta$ and $C_{abaa}=-C_{abbb}=C_{aaab}=-C_{bbab}=\Lambda_2(\alpha_{11}\alpha_{12}-\alpha_{22}\alpha_{21})=\Lambda_2\sin(2\,\theta)$, see \cite{Fuentes:Barberis-Blostein:2007, Barberis-Blostein-Fuentes-Schuller:2008,Sabin:Barberis-Blostein:Fuentes:2014}. Here $\Lambda_2$ is a free parameter corresponding to the two-body interaction strength. The term $\hat{H}_\textrm{cte}:=\hbar\,\Lambda_2\,(\hat{a}^\dag\,\hat{a}+\hat{b}^\dag\,\hat{b})$ commutes with the full Hamiltonian and can be ignored since it just adds a global phase to the time evolution of the state.

The Hamiltonian $\hat{H}=\hat{H}_{bs}+\hat{H}_2$ is an extension of the two-mode Bose-Hubbard \cite{Dalfovo:Giorgini:1999} and Lipkin-Meshkov-Glick \cite{Lipkin:Meshkov:1965,Paraoanu:Kohler:2001} models. It can be applied, for example, to describe the dynamics of the two-mode Bose-Einstein Condensate (BEC). The bi-quadratic interactions in $\hat{H}_2$ are elastic and mode-mixing collisions. In a double-well BEC, the operators   $\hat{a}^{\dag}\,\hat{a}^\dag\,\hat{a}\,\hat{a}$ and $\hat{b}^{\dag}\,\hat{b}^\dag\,\hat{b}\,\hat{b}$ are on-site elastic collisions and $\hat{a}^{\dag}\,\hat{b}^\dag\,\hat{a}\,\hat{b}$ are elastic collisions at the wave-function overlap. Mode-changing interactions are $\hat{a}^{\dag}\,\hat{b}^\dag\,\hat{a}\,\hat{a}$, $\hat{a}^{\dag}\,\hat{b}^\dag\,\hat{b}\,\hat{b}$, $\hat{a}^{\dag}\,\hat{a}^\dag\,\hat{a}\,\hat{b}$ and $\hat{b}^{\dag}\,\hat{b}^\dag\,\hat{a}\,\hat{b}$, which correspond to collision assisted tunnelling, while $\hat{b}^{\dag}\,\hat{b}^\dag\,\hat{a}\,\hat{a}$ and $\hat{a}^{\dag}\,\hat{a}^\dag\,\hat{b}\,\hat{b}$ correspond to two-body coherent tunneling. The two-mode Bose-Einstein Condensate Hamiltonian $\hat{H}_{\textrm{BEC}}$ is obtained using a two-mode approximation of the wave-function of the form $\hat{\phi}_{\theta}=\phi_a \hat{a}+\phi_b \hat{b}$, for appropriate functions $\phi_a$ and $\phi_b$, and it reads
\begin{align}\label{operator:BEC2:phi}
\hat{H}_{\textrm{BEC}}=\frac{\hbar}{2}\int dx\,\hat{\phi}_{\theta}^\dag \,\hat{H}_t\hat{\phi}_{\theta}+\Lambda_2\int\hat{\phi}^{\dag}_{\theta} \,\hat{\phi}^\dag_{\theta}\hat{\phi}_{\theta}\hat{\phi}_{\theta},
\end{align}
 where $\hat{H}_t=\nabla^2+V(x)$ is the double well Hamiltonian. The functions $\phi_a$ and $\phi_b$ are related to the functions $\tilde{\phi}_a$ and $\tilde{\phi}_b$ for which the Hamiltonian becomes diagonal through $\phi_a=\cos\theta\,\tilde{\phi}_a+\sin\theta\tilde{\phi}_b$ and $\phi_b=-\sin\theta\,\tilde{\phi}_a+\cos\theta\tilde{\phi}_b$. All details can be found in the appropriate literature \cite{Sabin:Barberis-Blostein:Fuentes:2014}. Note that it is common in the mathematical treatment of this model to neglect assisted and two-body tunnelling terms, however, keeping these terms yields Hamiltonians with exact solutions \cite{Sabin:Barberis-Blostein:2015}. In the covariance matrix formalism, the Hamiltonian is given by
\begin{align}\label{operator:BEC2:phi}
\hat{H}_{\textrm{BEC}}=\frac{\hbar}{2}\left(\hat{\mathbb{X}}^\dag_{\theta} \,\tilde{\boldsymbol{\kappa}}\,\hat{\mathbb{X}}_{\theta}\right)+\Lambda_2\left(\hat{\mathbb{X}}^\dag_{\theta} \,\tilde{\boldsymbol{\kappa}}\,\hat{\mathbb{X}}_{\theta}\right)^2.
\end{align}
Collision terms are usually written in normal order for mathematical convenience. However, in the covariance matrix formalism, it is more convenient to write collision and higher order terms as powers of the quadratic Hamiltonian $\hat{\mathbb{X}}^\dag_{\theta} \,\tilde{\boldsymbol{\kappa}}\,\hat{\mathbb{X}}_{\theta}$.
Previous work \cite{Fuentes:Barberis-Blostein:2007} showed that the Hamiltonian can be written as
\begin{align}\label{BEC:model:Hamtiltonian}
\hat{H}_{\textrm{BEC}}=\hat{U}_{\theta}^\dag\,\left(\hat{H}_0+\Lambda_2\hat{H}_0^2\right)\,\hat{U}_{\theta},
\end{align}
where the operator $\hat{U}_{\theta}=e^{\frac{\theta}{2}(\hat{a}^\dag\hat{b}+\hat{b}^\dag\hat{a})}$ is the unitary representation of the symplectic matrix $\boldsymbol{s}_{\theta}$. Hamiltonians with n-order interactions take the form
\begin{align}\label{BEC:model:HamtiltonianN}
\hat{H}=\hat{U}_{\theta}^\dag\,\left(\sum_{p=1}^{N}\Lambda_p\hat{H}_0^p\right)\,\hat{U}_{\theta},
\end{align}
where $\Lambda_p$ is the strength of the $p$-order interaction term. 

As an example of the application of higher order Hamiltonians given by \eqref{BEC:model:HamtiltonianN}, a model including two-body $(p=2)$ and three-body $(p=3)$ elastic and mode-exchange collisions is solved using the techniques described above \cite{Sabin:Barberis-Blostein:2015}. This enabled the understanding of the effects of three-body collisions in a two-mode BEC given by sextic terms such as $\hat{a}^{\dag}\hat{a}^{\dag}\hat{a}^{\dag}\hat{a}\hat{a}\hat{a}$. The analysis shows that three-body collisions change the probability distribution of the ground state as well as the dynamics of the relative population. Interestingly,  three-body interactions can inhibit collapse of the relative population, an effect usually seen when two-body collisions are present.

In this paper we are interested developing methods to solve Hamiltonians that not only include mode-mixing terms but also squeezing terms. By using the symplectic matrix formalism we are able to extend the Hamitonian \eqref{BEC:model:HamtiltonianN} to include simultaneously squeezing and higher order interactions. Writing $n$-order interactions in terms of a general symplectic transformation $\boldsymbol{s}$ we have $\hat{H}_n=\Lambda_n\frac{\hbar}{2}(\hat{\mathbb{X}_{\textrm{s}}}^\dag\,\tilde{\boldsymbol{\kappa}}\,\hat{\mathbb{X}}_{\textrm{s}})^n$, where $\hat{\mathbb{X}}_{\textrm{s}}=\boldsymbol{s}\hat{\mathbb{X}}$.
This expression now has non-vanishing $\beta$ coefficients. The diagonal Hamiltonian $H_0$ now includes two-mode squeezing terms $\hat{a}^{\dag}\hat{a}^{\dag}+\hat{a}\hat{a}$, $\hat{b}^{\dag}\hat{b}^{\dag}+\hat{b}\hat{b}$ and $\hat{a}^{\dag}\hat{b}^{\dag}+\hat{a}\hat{b}$, while the bi-quadratic Hamiltonian contains higher order squeezing terms such as $\hat{a}^{\dag}\hat{b}^{\dag}\hat{a}^{\dag}\hat{b}^{\dag}+\hat{a}\hat{b}\hat{a}\hat{b}$. The Hamiltonian that includes mode-mixing and squeezing interactions up to order $N$ reads
\begin{align}
\hat{H}_N=\frac{\hbar}{2}\,\sum_{p=1}^N\Lambda_p\left(\hat{\mathbb{X}}_{\textrm{s}}^\dag\tilde{\boldsymbol{\kappa}}\,\hat{\mathbb{X}}_{\textrm{s}}\right)^p,
\end{align} 
and it is diagonalizable by solving as a function of the interaction strengths $\Lambda_p$ and the Bogoliubov coefficients contained in the matrix $\boldsymbol{s}$. Applications, which we will present in a followup paper, enable finding analytical solutions for squeezing in the two-mode BEC and higher order squeezing in parametric down-conversion.

\section{Circuit implementation of the ultrastrong coupling time evolution}\label{circuit:section}
Here we show that the time evolution induced by the ultrastrong coupling Hamiltonian can be decomposed into a sequence of simpler operations, which we interpret as providing a simple way to implement a two-mode bosonic quantum channel. Detailed calculations can be found in Appendix \ref{channel:decomposition:appendix}.

\subsection{Circuit decomposition}

The foundation for this decomposition is the important result that states that any linear unitary operator of $N$ bosonic modes can be decomposed as a sequence of three operations: i) a generalized beam splitter of $N$ modes; ii) single mode squeezing of each mode; iii) another generilzed beam splitter of $N$ modes (in general different from the first one) \cite{Braunstein:2005}.

In our language, this means that the time evolution symplectic matrix \eqref{symplectic:expression:strong:coupling} can be written as 
\begin{align}\label{evolution:channel:decomposition}
\boldsymbol{S}(t)\equiv\boldsymbol{o}(-\phi) \boldsymbol{s}_\text{q}(-\underline{r})\boldsymbol{o}(-\varphi) e^{\boldsymbol{\Omega} \tilde{\boldsymbol{\kappa}} t} \boldsymbol{o}(\varphi) \boldsymbol{s}_\text{q}(\underline{r}) \boldsymbol{o}(\phi).
\end{align}
Here, $\boldsymbol{o}(\varphi)$, $\boldsymbol{s}_\text{q}(\underline{r})$ and $\boldsymbol{o}(\phi)$ are symplectic matrices in their own right, which represent the generalized beam splitter ($\boldsymbol{o}(\varphi)$ and $\boldsymbol{o}(\phi)$), and the squeezer ($\boldsymbol{s}_\text{q}(\underline{r})$). Or, in the language of quantum optics, the matrix $\boldsymbol{o}(\theta)$ implements a mode mixing channel, while the matrix $\boldsymbol{s}_\text{q}(\underline{r})$ implements a single mode squeezing channel on both modes. We have also collected the squeezing parameters $r_\text{a},r_\text{b}$ in the vector $\underline{r}:=(r_\text{a},r_\text{b})$.

To obtain our decomposition \eqref{evolution:channel:decomposition} we have used the expression $\boldsymbol{S}(t)=\boldsymbol{s}^{-1}\,\exp[\boldsymbol{\Omega}\,\tilde{\boldsymbol{\kappa}}\,t]\,\boldsymbol{s}$ obtained in \eqref{main:expression:symplectic:matrix}, we have applied the aforementioned result to the symplectic matrix $\boldsymbol{s}$, and we have used the following algebraic properties: the matrix $\boldsymbol{o}(\psi)$ has real valued entries, $\boldsymbol{o}(\psi)\,\boldsymbol{o}^\text{Tp}(\psi)=\boldsymbol{o}(\psi)\,\boldsymbol{o}(-\psi)=\mathds{1}$, $\boldsymbol{s}_\text{q}^\dag(\underline{r})=\boldsymbol{s}_\text{q}(\underline{r})$ and $\boldsymbol{s}_\text{q}(-\underline{r})\boldsymbol{s}_\text{q}(\underline{r})=\mathds{1}$.
The expressions for the $4\times4$ orthogonal matrix $\boldsymbol{o}(\psi)$ and the $4\times4$ squeezing matrix $\boldsymbol{s}_\text{q}(\underline{r})$ can be found in \eqref{decomposition:explicit:expression:appendix}. Finally, recall that $\exp[\boldsymbol{\Omega} \tilde{\boldsymbol{\kappa}} t]=\text{diag}(e^{-i\kappa_+t},e^{-i\kappa_-t},e^{i\kappa_+t},e^{i\kappa_-t})$.

In our case, the angles $\phi$ and $\varphi$, and the squeezing parameters $r_\text{a}$ and $r_\text{b}$ are related to the original parameters of the problem through the definitions
\begin{align}\label{parameter:relations}
\tan(2\,\phi):=&2\frac{\alpha_{11}\,\alpha_{12}+\alpha_{21}\,\alpha_{22}}{2\alpha_{21}^2-2\alpha_{12}^2+\beta_{11}^2+\beta_{12}^2-\beta_{21}^2-\beta_{22}^2}\nonumber\\
\tan(2\,\varphi):=&2\frac{\beta_{11}\,\beta_{21}+\beta_{12}\,\beta_{22}}{\beta_{11}^2+\beta_{12}^2-\beta_{21}^2-\beta_{22}^2}\nonumber\\
\sinh^2(r_\text{a}):=&\frac{1}{2}\left[\beta_{11}^2+\beta_{12}^2+\beta_{21}^2+\beta_{22}^2+K\right]\nonumber\\
\sinh^2(r_\text{b}):=&\frac{1}{2}\left[\beta_{11}^2+\beta_{12}^2+\beta_{21}^2+\beta_{22}^2-K\right].
\end{align}
Here, the expressions for the coefficients $\alpha_{nm}$ are given in \eqref{alpha:beta:expressions} and we have defined $K^2:=(\beta_{11}^2-\beta_{22}^2)^2+(\beta_{12}^2-\beta_{21}^2)^2+2(\beta_{11}^2+\beta_{22}^2)(\beta_{12}^2+\beta_{21}^2)+4\beta_{11}\beta_{12}\beta_{21}\beta_{22}$ for ease of presentation. Importantly, the squeezing in the evolution, which is quantified by $r_\text{a}$ and $r_\text{b}$, vanishes if \textit{all} beta-coefficients $\beta_{nm}$ vanish. Therefore, the presence of  $\beta_{nm}$ coefficients is necessary for the existence of squeezing in the system.

\begin{figure}[ht!]
    \includegraphics[width=\linewidth]{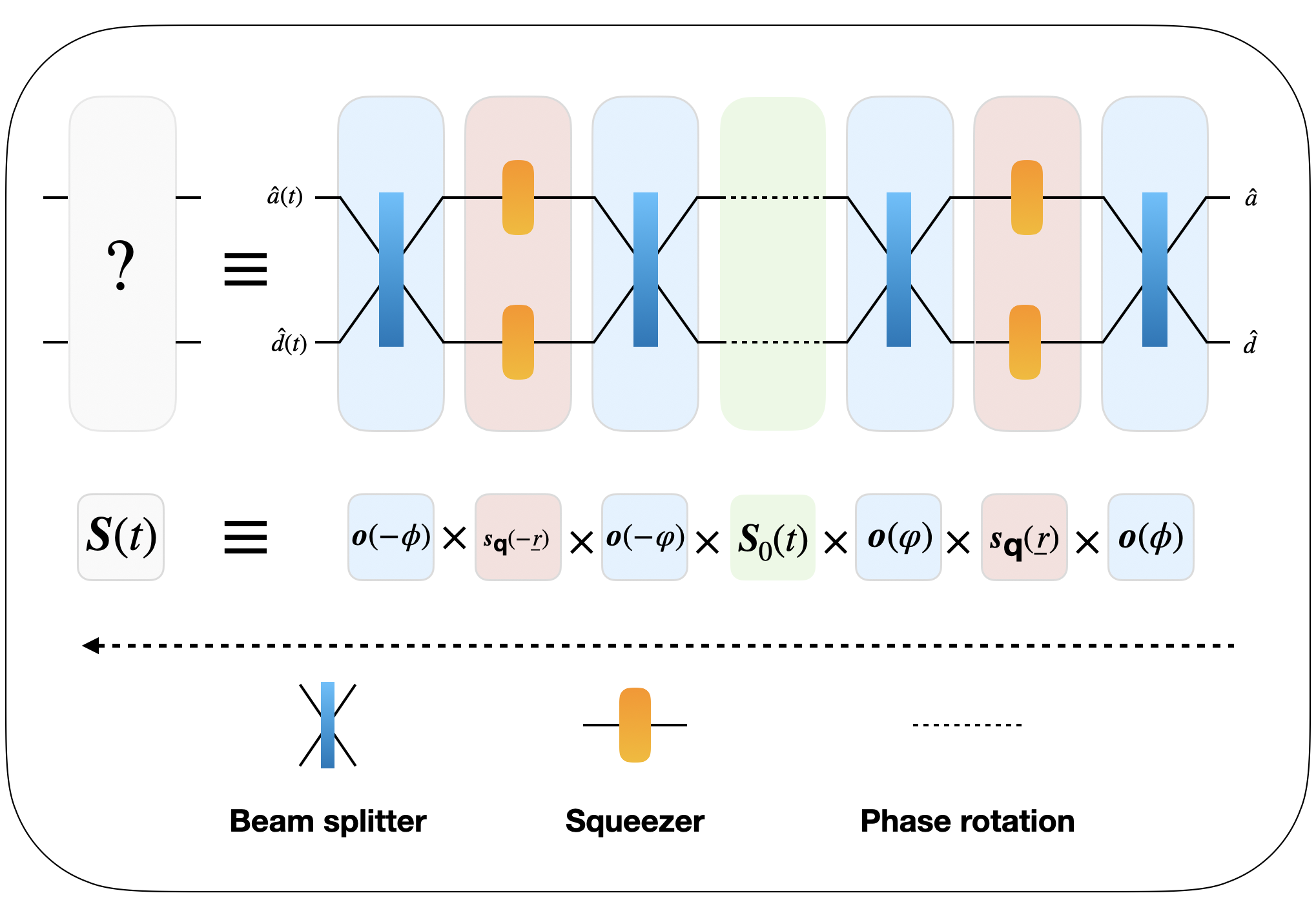}
    \caption{\label{figure:one}Decomposition of the total time evolution operator of modes $\hat{a}$ and $\hat{b}$, represented by the symplectic matrix $\boldsymbol{S}(t)$, in terms of simpler quantum optical operations. According to \eqref{evolution:channel:decomposition}, the time evolution channel can be constructed by a sequence of the following operations (from right to left): mixing of the modes with angle $\phi$; individual squeezing of each mode with parameters $\underline{r}=(r_\text{a},r_\text{b})$; mixing of the modes with angle $\varphi$; free evolution for time $t$ of the resulting modes with frequencies $\kappa_\pm$ respectively, implemented by $\boldsymbol{S}_0(t):=\exp[\boldsymbol{\Omega} \tilde{\boldsymbol{\kappa}} t]$; mixing of the modes with angle $-\varphi$; individual squeezing of each mode with parameter $-\underline{r}$; mixing of the modes with angle $-\phi$. Clearly, if the frequencies $\kappa_\pm$ are such that $\boldsymbol{S}_0(t)=\mathds{1}$ at some time $\tilde{t}$, then $\boldsymbol{S}(\tilde{t})=\mathds{1}$ as well.}
\end{figure}

Finally, in \autoref{figure:one} we give a pictorial representation of the operation \eqref{evolution:channel:decomposition} as a sequence of appropriate operations.

\subsection{Circuit implementation: considerations}
Let us now make a few considerations about the decomposed solution \eqref{evolution:channel:decomposition}. We start by noting that, when $e^{\boldsymbol{\Omega} \tilde{\boldsymbol{\kappa}} t}=\mathds{1}$, it immediately follows that $\boldsymbol{S}(t)\equiv\mathds{1}$. This means that the channel acts as the identity for times $\tilde{t}_n$ that satisfy the condition $(\kappa_+\pm\kappa_-)\tilde{t}_n=2\,n\,\pi$. The advantage of this channel decomposition picture is that it gives a clear understanding of the action of the time evolution on the creation and annihilation operators. In particular, it informs us that the time evolution can be essentially implemented by applying to the system some specific time independent quantum optical transformations (i.e., squeezing and mode mixing implemented by the matrices $\boldsymbol{s}_\text{q}(\underline{r})$ and $\boldsymbol{o}(\psi)$  respectively), while the evolution in time is provided solely by the free term $e^{\boldsymbol{\Omega} \tilde{\boldsymbol{\kappa}} t}$ with (symplectic) frequencies $\kappa_\pm$. 

Overall, the total action of the channel \eqref{evolution:channel:decomposition} is to two-mode squeeze the initial modes, with the addition of local mixing of the operators as well.

\subsection{Circuit implementation: applications}
Our work can be applied to any system that is modelled, at least in some regime, by the Hamiltonian \eqref{basic:hamiltonian}.

\begin{figure}[h!]
    \includegraphics[width=\linewidth]{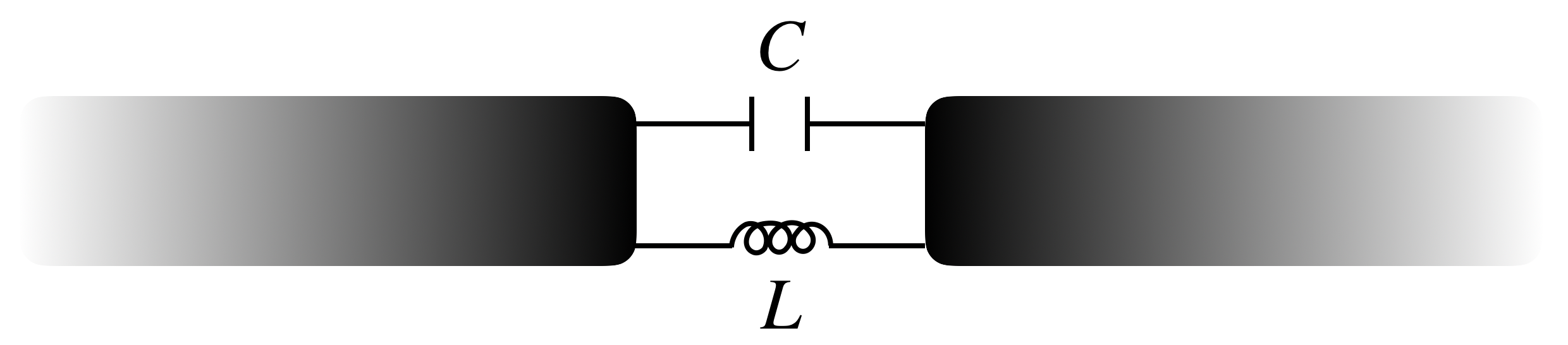}
    \caption{\label{figure:two}Microwave $LC$ resonator, where $L$ is the inductivity and $C$ the capacity. Such a resonator can be described using equations \eqref{second:basic:hamiltonian:physical} and \eqref{circuit:couplings}.}
\end{figure}

A particular implementation can be that of microwave circuits \cite{Paraoanu:2014,Wendin:2017,Krantz:Kjaergaard:2019} for circuit QED.
Circuit QED enables the most straightforward realization of the ultrastrong coupling regime, since the interaction between the electrical components can be engineered appropriately. So far, a lot of interest has been devoted to reach this regime with qubits coupled to resonators. There have been three main approaches: i) Observing the effect of Bloch-Siegert shift in driven systems \cite{Tuorila:Silveri:2010,Pietikainen:Danilin:2017,Pietikainen:Danilin:2018}; ii) Simulating the utrastrong coupling using additional fields and examining the effective Hamiltonians in a suitably-defined rotating frame \cite{Langford:Sagasrizabal:2017,Braumueller:Marthaler:2017,Li:Silveri:2013}; iii) Designing couplers that achieve a coupling strengths comparable with the frequency \cite{FronDiaz:Garcia-Ripoll:2017,Bosman:Gely:2017}. Here we take the latter approach and propose a circuit consisting of two LC oscillators with both capacitive and inductive coupling, realizing independently the two interaction terms studied before. In addition, each oscillator can be parametrically pumped, which will add the squeezing term. A very schematic depiction of the basic building block of such microwaves systems can be found in \autoref{figure:two}.

It is possible to engineer the interacting part $\hat{H}_\text{I}$ of the total Hamiltonian $\hat{H}=\hat{H}_0+\hat{H}_\text{I}$ as
\begin{align}\label{circuit:couplings}
\hat{H}_\text{I}=&g_\text{C}\,(\hat{a}-\hat{a}^\dag)\,(\hat{b}-\hat{b}^\dag) &&\text{Capacitive coupling}\nonumber\\
\hat{H}_\text{I}=&g_\text{L}\,(\hat{a}^\dag+\hat{a})\,(\hat{b}^\dag+\hat{b}) &&\text{Inductive coupling}
\end{align}
A detailed derivation of how to obtain such couplings can be found in Appendix \ref{superconducting:circuits:appendix}.

In this case it is easy to see that one can map such system to the one studied in this work by setting $g_\text{bs}=g_\text{sq}=g_\text{C}$ for the capacitive coupling, while setting $g_\text{bs}=-g_\text{sq}=g_\text{L}$ for the inductive coupling. Therefore, this implies that using a suitable combination of both capacitive and inductive couplings might allow to implement the Hamiltonian \eqref{basic:hamiltonian} with microwave systems~\cite{Wallraff:Schuster:2004}.

\section{Conclusions}
We have found the analytical solution of the time evolution of two harmonic oscillators interacting through a quadratic Hamiltonian with arbitrary parameters. As an important case of interest, we applied this result to obtain the full evolution of the oscillators in the ultrastrong coupling regime including single mode squeezing of each mode, where we provided all quantities as explicit functions of the parameters of the Hamiltonian.

Our solutions allow for the explicit evaluation of any quantity of interest. We have focussed on computing the average number of excitations in the system and the entropy of entanglement generated between the two oscillators. The latter, together with other measures of entanglement, can provide useful characterizations of the entanglement of the system for potential use in quantum information tasks \cite{quantuminfo}.  In addition, we found the existence of a second order phase transition, which is closely related to Dicke-like models operating in the thermodynamical limit. To characterize the transition, we computed the relevant defining parameters, such as the dynamical critical exponent. We also discussed how our techniques can be used to study particular families of Hamiltonians of higher order, namely, those that contain terms that are at least cubic in the difference between the quadrature operators of the oscillators. Furthermore, we were able to show that the time evolution can be decomposed as a sequence of simple, time-independent quantum optical operations and free evolution. This result illustrates the importance of the techniques used here, and the fact that complicated Hamiltonians can be implemented by sequences of simpler operations, and simulated efficiently.

Concluding, not only we provided an analytical solution to the time evolution of two harmonic oscillators interacting in the ultrastrong coupling regime, but we also provided a systematic way to implement such evolution through simple quantum optical operations within a simulation or in the laboratory. We leave it to further work to exploit these results for specific applications.

\acknowledgments
We thank Pablo Tieben and Anna Okopi\'nska for useful comments and suggestions. G.S.P. acknowledges support from Foundational Questions Institute (FQXi), from the European Commission project QUARTET (grant  agreement  no. 862644, FET  Open, Horizon 2020), and from the Academy of Finland through project 328193 and through the ``Finnish Center of Excellence in Quantum Technology QTF'' project 312296. D.E.B. acknowledges the Central European Institute of Technology (CEITEC) Nano RI for partial support. A.W.S. was partly funded by the Deutsche Forschungsgemeinschaft (DFG, German Research Foundation) under Germany’s Excellence Strategy – EXC-2123 QuantumFrontiers – 390837967.
This study was inspired by pioneering work of L. M. Narducci and collaborators.

\bibliographystyle{apsrev4-1}
\bibliography{NarducciSolutionBib}

\begin{thebibliography}{51}%
\makeatletter
\providecommand \@ifxundefined [1]{%
 \@ifx{#1\undefined}
}%
\providecommand \@ifnum [1]{%
 \ifnum #1\expandafter \@firstoftwo
 \else \expandafter \@secondoftwo
 \fi
}%
\providecommand \@ifx [1]{%
 \ifx #1\expandafter \@firstoftwo
 \else \expandafter \@secondoftwo
 \fi
}%
\providecommand \natexlab [1]{#1}%
\providecommand \enquote  [1]{``#1''}%
\providecommand \bibnamefont  [1]{#1}%
\providecommand \bibfnamefont [1]{#1}%
\providecommand \citenamefont [1]{#1}%
\providecommand \href@noop [0]{\@secondoftwo}%
\providecommand \href [0]{\begingroup \@sanitize@url \@href}%
\providecommand \@href[1]{\@@startlink{#1}\@@href}%
\providecommand \@@href[1]{\endgroup#1\@@endlink}%
\providecommand \@sanitize@url [0]{\catcode `\\12\catcode `\$12\catcode
  `\&12\catcode `\#12\catcode `\^12\catcode `\_12\catcode `\%12\relax}%
\providecommand \@@startlink[1]{}%
\providecommand \@@endlink[0]{}%
\providecommand \url  [0]{\begingroup\@sanitize@url \@url }%
\providecommand \@url [1]{\endgroup\@href {#1}{\urlprefix }}%
\providecommand \urlprefix  [0]{URL }%
\providecommand \Eprint [0]{\href }%
\providecommand \doibase [0]{http://dx.doi.org/}%
\providecommand \selectlanguage [0]{\@gobble}%
\providecommand \bibinfo  [0]{\@secondoftwo}%
\providecommand \bibfield  [0]{\@secondoftwo}%
\providecommand \translation [1]{[#1]}%
\providecommand \BibitemOpen [0]{}%
\providecommand \bibitemStop [0]{}%
\providecommand \bibitemNoStop [0]{.\EOS\space}%
\providecommand \EOS [0]{\spacefactor3000\relax}%
\providecommand \BibitemShut  [1]{\csname bibitem#1\endcsname}%
\let\auto@bib@innerbib\@empty
\bibitem [{\citenamefont {Aspelmeyer}\ \emph {et~al.}(2014)\citenamefont
  {Aspelmeyer}, \citenamefont {Kippenberg},\ and\ \citenamefont
  {Marquardt}}]{Aspelmeyer:Kippenberg:2014}%
  \BibitemOpen
  \bibfield  {author} {\bibinfo {author} {\bibfnamefont {M.}~\bibnamefont
  {Aspelmeyer}}, \bibinfo {author} {\bibfnamefont {T.~J.}\ \bibnamefont
  {Kippenberg}}, \ and\ \bibinfo {author} {\bibfnamefont {F.}~\bibnamefont
  {Marquardt}},\ }\href@noop {} {\bibfield  {journal} {\bibinfo  {journal}
  {Rev. Mod. Phys.}\ }\textbf {\bibinfo {volume} {86}},\ \bibinfo {pages}
  {1391} (\bibinfo {year} {2014})}\BibitemShut {NoStop}%
\bibitem [{\citenamefont {Dalfovo}\ \emph {et~al.}(1999)\citenamefont
  {Dalfovo}, \citenamefont {Giorgini}, \citenamefont {Pitaevskii},\ and\
  \citenamefont {Stringari}}]{Dalfovo:Giorgini:1999}%
  \BibitemOpen
  \bibfield  {author} {\bibinfo {author} {\bibfnamefont {F.}~\bibnamefont
  {Dalfovo}}, \bibinfo {author} {\bibfnamefont {S.}~\bibnamefont {Giorgini}},
  \bibinfo {author} {\bibfnamefont {L.~P.}\ \bibnamefont {Pitaevskii}}, \ and\
  \bibinfo {author} {\bibfnamefont {S.}~\bibnamefont {Stringari}},\ }\href@noop
  {} {\bibfield  {journal} {\bibinfo  {journal} {Rev. Mod. Phys.}\ }\textbf
  {\bibinfo {volume} {71}},\ \bibinfo {pages} {463} (\bibinfo {year}
  {1999})}\BibitemShut {NoStop}%
\bibitem [{\citenamefont {Blais}\ \emph {et~al.}(2004)\citenamefont {Blais},
  \citenamefont {Huang}, \citenamefont {Wallraff}, \citenamefont {Girvin},\
  and\ \citenamefont {Schoelkopf}}]{Blais:Huang:2004}%
  \BibitemOpen
  \bibfield  {author} {\bibinfo {author} {\bibfnamefont {A.}~\bibnamefont
  {Blais}}, \bibinfo {author} {\bibfnamefont {R.-S.}\ \bibnamefont {Huang}},
  \bibinfo {author} {\bibfnamefont {A.}~\bibnamefont {Wallraff}}, \bibinfo
  {author} {\bibfnamefont {S.~M.}\ \bibnamefont {Girvin}}, \ and\ \bibinfo
  {author} {\bibfnamefont {R.~J.}\ \bibnamefont {Schoelkopf}},\ }\href
  {\doibase 10.1103/PhysRevA.69.062320} {\bibfield  {journal} {\bibinfo
  {journal} {Phys. Rev. A}\ }\textbf {\bibinfo {volume} {69}},\ \bibinfo
  {pages} {062320} (\bibinfo {year} {2004})}\BibitemShut {NoStop}%
\bibitem [{\citenamefont {Emary}\ and\ \citenamefont
  {Brandes}(2003{\natexlab{a}})}]{Emary:Brandes:2003}%
  \BibitemOpen
  \bibfield  {author} {\bibinfo {author} {\bibfnamefont {C.}~\bibnamefont
  {Emary}}\ and\ \bibinfo {author} {\bibfnamefont {T.}~\bibnamefont
  {Brandes}},\ }\href {\doibase 10.1103/PhysRevE.67.066203} {\bibfield
  {journal} {\bibinfo  {journal} {Phys. Rev. E}\ }\textbf {\bibinfo {volume}
  {67}},\ \bibinfo {pages} {066203} (\bibinfo {year}
  {2003}{\natexlab{a}})}\BibitemShut {NoStop}%
\bibitem [{\citenamefont {Emary}\ and\ \citenamefont
  {Brandes}(2003{\natexlab{b}})}]{Emary:Brandes:2003:v2}%
  \BibitemOpen
  \bibfield  {author} {\bibinfo {author} {\bibfnamefont {C.}~\bibnamefont
  {Emary}}\ and\ \bibinfo {author} {\bibfnamefont {T.}~\bibnamefont
  {Brandes}},\ }\href {\doibase 10.1103/PhysRevLett.90.044101} {\bibfield
  {journal} {\bibinfo  {journal} {Phys. Rev. Lett.}\ }\textbf {\bibinfo
  {volume} {90}},\ \bibinfo {pages} {044101} (\bibinfo {year}
  {2003}{\natexlab{b}})}\BibitemShut {NoStop}%
\bibitem [{\citenamefont {Cuccoli}\ \emph {et~al.}(1995)\citenamefont
  {Cuccoli}, \citenamefont {Giachetti}, \citenamefont {Tognetti}, \citenamefont
  {Vaia},\ and\ \citenamefont {Verrucchi}}]{Cuccoli:Giachetti:1995}%
  \BibitemOpen
  \bibfield  {author} {\bibinfo {author} {\bibfnamefont {A.}~\bibnamefont
  {Cuccoli}}, \bibinfo {author} {\bibfnamefont {R.}~\bibnamefont {Giachetti}},
  \bibinfo {author} {\bibfnamefont {V.}~\bibnamefont {Tognetti}}, \bibinfo
  {author} {\bibfnamefont {R.}~\bibnamefont {Vaia}}, \ and\ \bibinfo {author}
  {\bibfnamefont {P.}~\bibnamefont {Verrucchi}},\ }\href {\doibase
  10.1088/0953-8984/7/41/003} {\bibfield  {journal} {\bibinfo  {journal}
  {Journal of Physics: Condensed Matter}\ }\textbf {\bibinfo {volume} {7}},\
  \bibinfo {pages} {7891} (\bibinfo {year} {1995})}\BibitemShut {NoStop}%
\bibitem [{\citenamefont {Haroche}\ and\ \citenamefont
  {Raimond}(2010)}]{Haroche:Raimond:2010}%
  \BibitemOpen
  \bibfield  {author} {\bibinfo {author} {\bibfnamefont {S.}~\bibnamefont
  {Haroche}}\ and\ \bibinfo {author} {\bibfnamefont {J.-M.}\ \bibnamefont
  {Raimond}},\ }\href@noop {} {\emph {\bibinfo {title} {{Exploring the Quantum:
  Atoms, Cavities, and Photons}}}},\ {Oxford Graduate texts}\ (\bibinfo
  {publisher} {Oxford University Press},\ \bibinfo {year} {2010})\BibitemShut
  {NoStop}%
\bibitem [{\citenamefont {Estes}\ \emph {et~al.}(1968)\citenamefont {Estes},
  \citenamefont {Keil},\ and\ \citenamefont {Narducci}}]{Estes:Keil:1968}%
  \BibitemOpen
  \bibfield  {author} {\bibinfo {author} {\bibfnamefont {L.~E.}\ \bibnamefont
  {Estes}}, \bibinfo {author} {\bibfnamefont {T.~H.}\ \bibnamefont {Keil}}, \
  and\ \bibinfo {author} {\bibfnamefont {L.~M.}\ \bibnamefont {Narducci}},\
  }\href@noop {} {\bibfield  {journal} {\bibinfo  {journal} {Phys. Rev.}\
  }\textbf {\bibinfo {volume} {175}},\ \bibinfo {pages} {286} (\bibinfo {year}
  {1968})}\BibitemShut {NoStop}%
\bibitem [{\citenamefont {Dippel}\ \emph {et~al.}(1994)\citenamefont {Dippel},
  \citenamefont {Schmelcher},\ and\ \citenamefont
  {Cederbaum}}]{Dippel:Schmelcher:1994}%
  \BibitemOpen
  \bibfield  {author} {\bibinfo {author} {\bibfnamefont {O.}~\bibnamefont
  {Dippel}}, \bibinfo {author} {\bibfnamefont {P.}~\bibnamefont {Schmelcher}},
  \ and\ \bibinfo {author} {\bibfnamefont {L.~S.}\ \bibnamefont {Cederbaum}},\
  }\href {\doibase 10.1103/PhysRevA.49.4415} {\bibfield  {journal} {\bibinfo
  {journal} {Phys. Rev. A}\ }\textbf {\bibinfo {volume} {49}},\ \bibinfo
  {pages} {4415} (\bibinfo {year} {1994})}\BibitemShut {NoStop}%
\bibitem [{\citenamefont {Mukhopadhyay}(2018)}]{Mukhopadhyay:2018}%
  \BibitemOpen
  \bibfield  {author} {\bibinfo {author} {\bibfnamefont {A.}~\bibnamefont
  {Mukhopadhyay}},\ }\href@noop {} {\bibfield  {journal} {\bibinfo  {journal}
  {Journal of Physics Communications}\ }\textbf {\bibinfo {volume} {2}},\
  \bibinfo {pages} {055004} (\bibinfo {year} {2018})}\BibitemShut {NoStop}%
\bibitem [{\citenamefont {Bruschi}\ \emph {et~al.}(2013)\citenamefont
  {Bruschi}, \citenamefont {Lee},\ and\ \citenamefont
  {Fuentes}}]{Bruschi:Lee:2013}%
  \BibitemOpen
  \bibfield  {author} {\bibinfo {author} {\bibfnamefont {D.~E.}\ \bibnamefont
  {Bruschi}}, \bibinfo {author} {\bibfnamefont {A.~R.}\ \bibnamefont {Lee}}, \
  and\ \bibinfo {author} {\bibfnamefont {I.}~\bibnamefont {Fuentes}},\
  }\href@noop {} {\bibfield  {journal} {\bibinfo  {journal} {Journal of Physics
  A: Mathematical and Theoretical}\ }\textbf {\bibinfo {volume} {46}},\
  \bibinfo {pages} {165303} (\bibinfo {year} {2013})}\BibitemShut {NoStop}%
\bibitem [{\citenamefont {Sandoval-Santana}\ \emph {et~al.}(2016)\citenamefont
  {Sandoval-Santana}, \citenamefont {Ibarra-Sierra}, \citenamefont {Cardoso},\
  and\ \citenamefont {Kunold}}]{Sandoval-Santana:Ibarra-Sierra:2016}%
  \BibitemOpen
  \bibfield  {author} {\bibinfo {author} {\bibfnamefont {J.~C.}\ \bibnamefont
  {Sandoval-Santana}}, \bibinfo {author} {\bibfnamefont {V.~G.}\ \bibnamefont
  {Ibarra-Sierra}}, \bibinfo {author} {\bibfnamefont {J.~L.}\ \bibnamefont
  {Cardoso}}, \ and\ \bibinfo {author} {\bibfnamefont {A.}~\bibnamefont
  {Kunold}},\ }\href@noop {} {\bibfield  {journal} {\bibinfo  {journal}
  {Journal of Mathematical Physics}\ }\textbf {\bibinfo {volume} {57}},\
  \bibinfo {pages} {042104} (\bibinfo {year} {2016})}\BibitemShut {NoStop}%
\bibitem [{\citenamefont {Baksic}\ and\ \citenamefont
  {Ciuti}(2014)}]{Baksic:Ciuti:2014}%
  \BibitemOpen
  \bibfield  {author} {\bibinfo {author} {\bibfnamefont {A.}~\bibnamefont
  {Baksic}}\ and\ \bibinfo {author} {\bibfnamefont {C.}~\bibnamefont {Ciuti}},\
  }\href {\doibase 10.1103/PhysRevLett.112.173601} {\bibfield  {journal}
  {\bibinfo  {journal} {Phys. Rev. Lett.}\ }\textbf {\bibinfo {volume} {112}},\
  \bibinfo {pages} {173601} (\bibinfo {year} {2014})}\BibitemShut {NoStop}%
\bibitem [{\citenamefont {Fuentes-Schuller}\ and\ \citenamefont
  {Barberis-Blostein}(2007)}]{Fuentes:Barberis-Blostein:2007}%
  \BibitemOpen
  \bibfield  {author} {\bibinfo {author} {\bibfnamefont {I.}~\bibnamefont
  {Fuentes-Schuller}}\ and\ \bibinfo {author} {\bibfnamefont {P.}~\bibnamefont
  {Barberis-Blostein}},\ }\href {\doibase 10.1088/1751-8113/40/27/f04}
  {\bibfield  {journal} {\bibinfo  {journal} {Journal of Physics A:
  Mathematical and Theoretical}\ }\textbf {\bibinfo {volume} {40}},\ \bibinfo
  {pages} {F601} (\bibinfo {year} {2007})}\BibitemShut {NoStop}%
\bibitem [{\citenamefont {Zamastil}\ \emph {et~al.}(2000)\citenamefont
  {Zamastil}, \citenamefont {\u{C}\'i\u{z}ek},\ and\ \citenamefont
  {Sk\'ala}}]{Zamastil:Cizek:2000}%
  \BibitemOpen
  \bibfield  {author} {\bibinfo {author} {\bibfnamefont {J.}~\bibnamefont
  {Zamastil}}, \bibinfo {author} {\bibfnamefont {J.}~\bibnamefont
  {\u{C}\'i\u{z}ek}}, \ and\ \bibinfo {author} {\bibfnamefont {L.}~\bibnamefont
  {Sk\'ala}},\ }\href {\doibase 10.1103/PhysRevLett.84.5683} {\bibfield
  {journal} {\bibinfo  {journal} {Phys. Rev. Lett.}\ }\textbf {\bibinfo
  {volume} {84}},\ \bibinfo {pages} {5683} (\bibinfo {year}
  {2000})}\BibitemShut {NoStop}%
\bibitem [{\citenamefont {Liverts}\ \emph {et~al.}(2006)\citenamefont
  {Liverts}, \citenamefont {Mandelzweig},\ and\ \citenamefont
  {Tabakin}}]{Liverts:Mandelzweig:2006}%
  \BibitemOpen
  \bibfield  {author} {\bibinfo {author} {\bibfnamefont {E.~Z.}\ \bibnamefont
  {Liverts}}, \bibinfo {author} {\bibfnamefont {V.~B.}\ \bibnamefont
  {Mandelzweig}}, \ and\ \bibinfo {author} {\bibfnamefont {F.}~\bibnamefont
  {Tabakin}},\ }\href {\doibase 10.1063/1.2209769} {\bibfield  {journal}
  {\bibinfo  {journal} {Journal of Mathematical Physics}\ }\textbf {\bibinfo
  {volume} {47}},\ \bibinfo {pages} {062109} (\bibinfo {year} {2006})},\
  \Eprint {http://arxiv.org/abs/https://doi.org/10.1063/1.2209769}
  {https://doi.org/10.1063/1.2209769} \BibitemShut {NoStop}%
\bibitem [{\citenamefont {Nielsen}\ and\ \citenamefont
  {Chuang}(2000)}]{quantuminfo}%
  \BibitemOpen
  \bibfield  {author} {\bibinfo {author} {\bibfnamefont {M.~A.}\ \bibnamefont
  {Nielsen}}\ and\ \bibinfo {author} {\bibfnamefont {I.~L.}\ \bibnamefont
  {Chuang}},\ }\href@noop {} {\emph {\bibinfo {title} {Quantum computation and
  quantum information}}}\ (\bibinfo  {publisher} {Cambridge University Press},\
  \bibinfo {year} {2000})\BibitemShut {NoStop}%
\bibitem [{\citenamefont {Frisk~Kockum}\ \emph {et~al.}(2019)\citenamefont
  {Frisk~Kockum}, \citenamefont {Miranowicz}, \citenamefont {De~Liberato},
  \citenamefont {Savasta},\ and\ \citenamefont
  {Nori}}]{Kockum:Miranowicz:2019}%
  \BibitemOpen
  \bibfield  {author} {\bibinfo {author} {\bibfnamefont {A.}~\bibnamefont
  {Frisk~Kockum}}, \bibinfo {author} {\bibfnamefont {A.}~\bibnamefont
  {Miranowicz}}, \bibinfo {author} {\bibfnamefont {S.}~\bibnamefont
  {De~Liberato}}, \bibinfo {author} {\bibfnamefont {S.}~\bibnamefont
  {Savasta}}, \ and\ \bibinfo {author} {\bibfnamefont {F.}~\bibnamefont
  {Nori}},\ }\href {\doibase 10.1038/s42254-018-0006-2} {\bibfield  {journal}
  {\bibinfo  {journal} {Nature Reviews Physics}\ }\textbf {\bibinfo {volume}
  {1}},\ \bibinfo {pages} {19} (\bibinfo {year} {2019})}\BibitemShut {NoStop}%
\bibitem [{\citenamefont {Forn-D\'{\i}az}\ \emph {et~al.}(2019)\citenamefont
  {Forn-D\'{\i}az}, \citenamefont {Lamata}, \citenamefont {Rico}, \citenamefont
  {Kono},\ and\ \citenamefont {Solano}}]{Diaz:Lamata:2019}%
  \BibitemOpen
  \bibfield  {author} {\bibinfo {author} {\bibfnamefont {P.}~\bibnamefont
  {Forn-D\'{\i}az}}, \bibinfo {author} {\bibfnamefont {L.}~\bibnamefont
  {Lamata}}, \bibinfo {author} {\bibfnamefont {E.}~\bibnamefont {Rico}},
  \bibinfo {author} {\bibfnamefont {J.}~\bibnamefont {Kono}}, \ and\ \bibinfo
  {author} {\bibfnamefont {E.}~\bibnamefont {Solano}},\ }\href {\doibase
  10.1103/RevModPhys.91.025005} {\bibfield  {journal} {\bibinfo  {journal}
  {Rev. Mod. Phys.}\ }\textbf {\bibinfo {volume} {91}},\ \bibinfo {pages}
  {025005} (\bibinfo {year} {2019})}\BibitemShut {NoStop}%
\bibitem [{\citenamefont {Monteiro}\ \emph {et~al.}(2013)\citenamefont
  {Monteiro}, \citenamefont {Millen}, \citenamefont {Pender}, \citenamefont
  {Marquardt}, \citenamefont {Chang},\ and\ \citenamefont
  {Barker}}]{Monteiro:Millen:2013}%
  \BibitemOpen
  \bibfield  {author} {\bibinfo {author} {\bibfnamefont {T.~S.}\ \bibnamefont
  {Monteiro}}, \bibinfo {author} {\bibfnamefont {J.}~\bibnamefont {Millen}},
  \bibinfo {author} {\bibfnamefont {G.~A.~T.}\ \bibnamefont {Pender}}, \bibinfo
  {author} {\bibfnamefont {F.}~\bibnamefont {Marquardt}}, \bibinfo {author}
  {\bibfnamefont {D.}~\bibnamefont {Chang}}, \ and\ \bibinfo {author}
  {\bibfnamefont {P.~F.}\ \bibnamefont {Barker}},\ }\href {\doibase
  10.1088/1367-2630/15/1/015001} {\bibfield  {journal} {\bibinfo  {journal}
  {New Journal of Physics}\ }\textbf {\bibinfo {volume} {15}},\ \bibinfo
  {pages} {015001} (\bibinfo {year} {2013})}\BibitemShut {NoStop}%
\bibitem [{\citenamefont {del Pino}\ \emph {et~al.}(2015)\citenamefont {del
  Pino}, \citenamefont {Feist},\ and\ \citenamefont
  {Garc\'ia-Vidal}}]{delPino:Feist:2015}%
  \BibitemOpen
  \bibfield  {author} {\bibinfo {author} {\bibfnamefont {J.}~\bibnamefont {del
  Pino}}, \bibinfo {author} {\bibfnamefont {J.}~\bibnamefont {Feist}}, \ and\
  \bibinfo {author} {\bibfnamefont {F.~J.}\ \bibnamefont {Garc\'ia-Vidal}},\
  }\href@noop {} {\bibfield  {journal} {\bibinfo  {journal} {New Journal of
  Physics}\ }\textbf {\bibinfo {volume} {17}},\ \bibinfo {pages} {053040}
  (\bibinfo {year} {2015})}\BibitemShut {NoStop}%
\bibitem [{\citenamefont {Jin}\ \emph {et~al.}(2013)\citenamefont {Jin},
  \citenamefont {Rossini}, \citenamefont {Fazio}, \citenamefont {Leib},\ and\
  \citenamefont {Hartmann}}]{Jin:Rossini:2013}%
  \BibitemOpen
  \bibfield  {author} {\bibinfo {author} {\bibfnamefont {J.}~\bibnamefont
  {Jin}}, \bibinfo {author} {\bibfnamefont {D.}~\bibnamefont {Rossini}},
  \bibinfo {author} {\bibfnamefont {R.}~\bibnamefont {Fazio}}, \bibinfo
  {author} {\bibfnamefont {M.}~\bibnamefont {Leib}}, \ and\ \bibinfo {author}
  {\bibfnamefont {M.~J.}\ \bibnamefont {Hartmann}},\ }\href@noop {} {\bibfield
  {journal} {\bibinfo  {journal} {Phys. Rev. Lett.}\ }\textbf {\bibinfo
  {volume} {110}},\ \bibinfo {pages} {163605} (\bibinfo {year}
  {2013})}\BibitemShut {NoStop}%
\bibitem [{\citenamefont {Jin}\ \emph {et~al.}(2014)\citenamefont {Jin},
  \citenamefont {Rossini}, \citenamefont {Leib}, \citenamefont {Hartmann},\
  and\ \citenamefont {Fazio}}]{Jin:Rossini:2014}%
  \BibitemOpen
  \bibfield  {author} {\bibinfo {author} {\bibfnamefont {J.}~\bibnamefont
  {Jin}}, \bibinfo {author} {\bibfnamefont {D.}~\bibnamefont {Rossini}},
  \bibinfo {author} {\bibfnamefont {M.}~\bibnamefont {Leib}}, \bibinfo {author}
  {\bibfnamefont {M.~J.}\ \bibnamefont {Hartmann}}, \ and\ \bibinfo {author}
  {\bibfnamefont {R.}~\bibnamefont {Fazio}},\ }\href@noop {} {\bibfield
  {journal} {\bibinfo  {journal} {Phys. Rev. A}\ }\textbf {\bibinfo {volume}
  {90}},\ \bibinfo {pages} {023827} (\bibinfo {year} {2014})}\BibitemShut
  {NoStop}%
\bibitem [{\citenamefont {Adesso}\ \emph {et~al.}(2014)\citenamefont {Adesso},
  \citenamefont {Ragy},\ and\ \citenamefont {Lee}}]{Adesso:Ragy:2014}%
  \BibitemOpen
  \bibfield  {author} {\bibinfo {author} {\bibfnamefont {G.}~\bibnamefont
  {Adesso}}, \bibinfo {author} {\bibfnamefont {S.}~\bibnamefont {Ragy}}, \ and\
  \bibinfo {author} {\bibfnamefont {A.~R.}\ \bibnamefont {Lee}},\ }\href@noop
  {} {\bibfield  {journal} {\bibinfo  {journal} {Open Systems \& Information
  Dynamics}\ }\textbf {\bibinfo {volume} {21}},\ \bibinfo {pages} {1440001}
  (\bibinfo {year} {2014})}\BibitemShut {NoStop}%
\bibitem [{\citenamefont {Brown}\ \emph {et~al.}(2013)\citenamefont {Brown},
  \citenamefont {Mart\'{\i}n-Mart\'{\i}nez}, \citenamefont {Menicucci},\ and\
  \citenamefont {Mann}}]{Brown:Matrin-Martinez:2013}%
  \BibitemOpen
  \bibfield  {author} {\bibinfo {author} {\bibfnamefont {E.~G.}\ \bibnamefont
  {Brown}}, \bibinfo {author} {\bibfnamefont {E.}~\bibnamefont
  {Mart\'{\i}n-Mart\'{\i}nez}}, \bibinfo {author} {\bibfnamefont {N.~C.}\
  \bibnamefont {Menicucci}}, \ and\ \bibinfo {author} {\bibfnamefont {R.~B.}\
  \bibnamefont {Mann}},\ }\href@noop {} {\bibfield  {journal} {\bibinfo
  {journal} {Phys. Rev. D}\ }\textbf {\bibinfo {volume} {87}},\ \bibinfo
  {pages} {084062} (\bibinfo {year} {2013})}\BibitemShut {NoStop}%
\bibitem [{\citenamefont {Bruschi}\ and\ \citenamefont
  {Xuereb}(2018)}]{Bruschi:Xuereb:2018}%
  \BibitemOpen
  \bibfield  {author} {\bibinfo {author} {\bibfnamefont {D.~E.}\ \bibnamefont
  {Bruschi}}\ and\ \bibinfo {author} {\bibfnamefont {A.}~\bibnamefont
  {Xuereb}},\ }\href {\doibase 10.1088/1367-2630/aaca27} {\bibfield  {journal}
  {\bibinfo  {journal} {New Journal of Physics}\ }\textbf {\bibinfo {volume}
  {20}},\ \bibinfo {pages} {065004} (\bibinfo {year} {2018})}\BibitemShut
  {NoStop}%
\bibitem [{\citenamefont {Williamson}(1936)}]{Williamson:1923}%
  \BibitemOpen
  \bibfield  {author} {\bibinfo {author} {\bibfnamefont {J.}~\bibnamefont
  {Williamson}},\ }\href@noop {} {\bibfield  {journal} {\bibinfo  {journal}
  {American Journal of Mathematics}\ }\textbf {\bibinfo {volume} {58}},\
  \bibinfo {pages} {141} (\bibinfo {year} {1936})}\BibitemShut {NoStop}%
\bibitem [{\citenamefont {Portes}\ \emph {et~al.}(2008)\citenamefont {Portes},
  \citenamefont {Rodrigues}, \citenamefont {Duarte},\ and\ \citenamefont
  {Baseia}}]{Portes:Rodrigues:2008}%
  \BibitemOpen
  \bibfield  {author} {\bibinfo {author} {\bibfnamefont {D.}~\bibnamefont
  {Portes}}, \bibinfo {author} {\bibfnamefont {H.}~\bibnamefont {Rodrigues}},
  \bibinfo {author} {\bibfnamefont {S.~B.}\ \bibnamefont {Duarte}}, \ and\
  \bibinfo {author} {\bibfnamefont {B.}~\bibnamefont {Baseia}},\ }\href@noop {}
  {\bibfield  {journal} {\bibinfo  {journal} {The European Physical Journal D}\
  }\textbf {\bibinfo {volume} {48}},\ \bibinfo {pages} {145} (\bibinfo {year}
  {2008})}\BibitemShut {NoStop}%
\bibitem [{\citenamefont {Urz{\'u}a}\ \emph {et~al.}(2019)\citenamefont
  {Urz{\'u}a}, \citenamefont {Ramos-Prieto}, \citenamefont {Soto-Eguibar},
  \citenamefont {Arriz{\'o}n},\ and\ \citenamefont
  {Moya-Cessa}}]{Urzua:Ramos-Prieto:2019}%
  \BibitemOpen
  \bibfield  {author} {\bibinfo {author} {\bibfnamefont {A.~R.}\ \bibnamefont
  {Urz{\'u}a}}, \bibinfo {author} {\bibfnamefont {I.}~\bibnamefont
  {Ramos-Prieto}}, \bibinfo {author} {\bibfnamefont {F.}~\bibnamefont
  {Soto-Eguibar}}, \bibinfo {author} {\bibfnamefont {V.}~\bibnamefont
  {Arriz{\'o}n}}, \ and\ \bibinfo {author} {\bibfnamefont {H.~M.}\ \bibnamefont
  {Moya-Cessa}},\ }\href {\doibase 10.1038/s41598-019-53024-5} {\bibfield
  {journal} {\bibinfo  {journal} {Scientific Reports}\ }\textbf {\bibinfo
  {volume} {9}},\ \bibinfo {pages} {16800} (\bibinfo {year}
  {2019})}\BibitemShut {NoStop}%
\bibitem [{\citenamefont {Abdalla}(1994)}]{Abdalla:1994}%
  \BibitemOpen
  \bibfield  {author} {\bibinfo {author} {\bibfnamefont {M.~S.}\ \bibnamefont
  {Abdalla}},\ }\href@noop {} {\bibfield  {journal} {\bibinfo  {journal} {Il
  Nuovo Cimento B (1971-1996)}\ }\textbf {\bibinfo {volume} {109}},\ \bibinfo
  {pages} {443} (\bibinfo {year} {1994})}\BibitemShut {NoStop}%
\bibitem [{\citenamefont {Peres}(1996)}]{Peres:1996}%
  \BibitemOpen
  \bibfield  {author} {\bibinfo {author} {\bibfnamefont {A.}~\bibnamefont
  {Peres}},\ }\href@noop {} {\bibfield  {journal} {\bibinfo  {journal} {Phys.
  Rev. Lett.}\ }\textbf {\bibinfo {volume} {77}},\ \bibinfo {pages} {1413}
  (\bibinfo {year} {1996})}\BibitemShut {NoStop}%
\bibitem [{\citenamefont {Adesso}\ and\ \citenamefont
  {Illuminati}(2007)}]{Adesso:Illuminati:2007}%
  \BibitemOpen
  \bibfield  {author} {\bibinfo {author} {\bibfnamefont {G.}~\bibnamefont
  {Adesso}}\ and\ \bibinfo {author} {\bibfnamefont {F.}~\bibnamefont
  {Illuminati}},\ }\href@noop {} {\bibfield  {journal} {\bibinfo  {journal}
  {Journal of Physics A: Mathematical and Theoretical}\ }\textbf {\bibinfo
  {volume} {40}},\ \bibinfo {pages} {7821} (\bibinfo {year}
  {2007})}\BibitemShut {NoStop}%
\bibitem [{\citenamefont {Dragan}\ and\ \citenamefont
  {Fuentes}(2011)}]{Dragan:Fuentes:2011}%
  \BibitemOpen
  \bibfield  {author} {\bibinfo {author} {\bibfnamefont {A.}~\bibnamefont
  {Dragan}}\ and\ \bibinfo {author} {\bibfnamefont {I.}~\bibnamefont
  {Fuentes}},\ }\href@noop {} {\enquote {\bibinfo {title} {Probing the
  spacetime structure of vacuum entanglement},}\ } (\bibinfo {year} {2011}),\
  \Eprint {http://arxiv.org/abs/arXiv:1105.1192} {arXiv:1105.1192} \BibitemShut
  {NoStop}%
\bibitem [{\citenamefont {Barberis-Blostein}\ and\ \citenamefont
  {Fuentes-Schuller}(2008)}]{Barberis-Blostein-Fuentes-Schuller:2008}%
  \BibitemOpen
  \bibfield  {author} {\bibinfo {author} {\bibfnamefont {P.}~\bibnamefont
  {Barberis-Blostein}}\ and\ \bibinfo {author} {\bibfnamefont {I.}~\bibnamefont
  {Fuentes-Schuller}},\ }\href {\doibase 10.1103/PhysRevA.78.013641} {\bibfield
   {journal} {\bibinfo  {journal} {Phys. Rev. A}\ }\textbf {\bibinfo {volume}
  {78}},\ \bibinfo {pages} {013641} (\bibinfo {year} {2008})}\BibitemShut
  {NoStop}%
\bibitem [{\citenamefont {Sab\'in}\ \emph {et~al.}(2014)\citenamefont
  {Sab\'in}, \citenamefont {Barberis-Blostein}, \citenamefont {Hernández},\
  and\ \citenamefont {Fuentes}}]{Sabin:Barberis-Blostein:Fuentes:2014}%
  \BibitemOpen
  \bibfield  {author} {\bibinfo {author} {\bibfnamefont {C.}~\bibnamefont
  {Sab\'in}}, \bibinfo {author} {\bibfnamefont {P.}~\bibnamefont
  {Barberis-Blostein}}, \bibinfo {author} {\bibfnamefont {C.}~\bibnamefont
  {Hernández}}, \ and\ \bibinfo {author} {\bibfnamefont {I.}~\bibnamefont
  {Fuentes}},\ }\href@noop {} {\enquote {\bibinfo {title} {Analytical solution
  of a double-well bose-einstein condensate},}\ } (\bibinfo {year} {2014}),\
  \Eprint {http://arxiv.org/abs/https://arxiv.org/abs/1406.4984}
  {https://arxiv.org/abs/1406.4984} \BibitemShut {NoStop}%
\bibitem [{\citenamefont {{Lipkin}}\ \emph {et~al.}(1965)\citenamefont
  {{Lipkin}}, \citenamefont {{Meshkov}},\ and\ \citenamefont
  {{Glick}}}]{Lipkin:Meshkov:1965}%
  \BibitemOpen
  \bibfield  {author} {\bibinfo {author} {\bibfnamefont {H.~J.}\ \bibnamefont
  {{Lipkin}}}, \bibinfo {author} {\bibfnamefont {N.}~\bibnamefont {{Meshkov}}},
  \ and\ \bibinfo {author} {\bibfnamefont {A.~J.}\ \bibnamefont {{Glick}}},\
  }\href {\doibase 10.1016/0029-5582(65)90862-X} {\bibfield  {journal}
  {\bibinfo  {journal} {Nuclear Physics}\ }\textbf {\bibinfo {volume} {62}},\
  \bibinfo {pages} {188} (\bibinfo {year} {1965})}\BibitemShut {NoStop}%
\bibitem [{\citenamefont {Paraoanu}\ \emph {et~al.}(2001)\citenamefont
  {Paraoanu}, \citenamefont {Kohler}, \citenamefont {Sols},\ and\ \citenamefont
  {Leggett}}]{Paraoanu:Kohler:2001}%
  \BibitemOpen
  \bibfield  {author} {\bibinfo {author} {\bibfnamefont {G.-S.}\ \bibnamefont
  {Paraoanu}}, \bibinfo {author} {\bibfnamefont {S.}~\bibnamefont {Kohler}},
  \bibinfo {author} {\bibfnamefont {F.}~\bibnamefont {Sols}}, \ and\ \bibinfo
  {author} {\bibfnamefont {A.~J.}\ \bibnamefont {Leggett}},\ }\href@noop {}
  {\bibfield  {journal} {\bibinfo  {journal} {Journal of Physics B: Atomic,
  Molecular and Optical Physics}\ }\textbf {\bibinfo {volume} {34}},\ \bibinfo
  {pages} {4689} (\bibinfo {year} {2001})}\BibitemShut {NoStop}%
\bibitem [{\citenamefont {Sab\'in}\ \emph {et~al.}(2015)\citenamefont
  {Sab\'in}, \citenamefont {Barberis-Blostein}, \citenamefont {Hernández},
  \citenamefont {Mann},\ and\ \citenamefont
  {Fuentes}}]{Sabin:Barberis-Blostein:2015}%
  \BibitemOpen
  \bibfield  {author} {\bibinfo {author} {\bibfnamefont {C.}~\bibnamefont
  {Sab\'in}}, \bibinfo {author} {\bibfnamefont {P.}~\bibnamefont
  {Barberis-Blostein}}, \bibinfo {author} {\bibfnamefont {C.}~\bibnamefont
  {Hernández}}, \bibinfo {author} {\bibfnamefont {R.~B.}\ \bibnamefont
  {Mann}}, \ and\ \bibinfo {author} {\bibfnamefont {I.}~\bibnamefont
  {Fuentes}},\ }\href {\doibase 10.1063/1.4936314} {\bibfield  {journal}
  {\bibinfo  {journal} {Journal of Mathematical Physics}\ }\textbf {\bibinfo
  {volume} {56}},\ \bibinfo {pages} {112102} (\bibinfo {year} {2015})},\
  \Eprint {http://arxiv.org/abs/https://doi.org/10.1063/1.4936314}
  {https://doi.org/10.1063/1.4936314} \BibitemShut {NoStop}%
\bibitem [{\citenamefont {Braunstein}(2005)}]{Braunstein:2005}%
  \BibitemOpen
  \bibfield  {author} {\bibinfo {author} {\bibfnamefont {S.~L.}\ \bibnamefont
  {Braunstein}},\ }\href@noop {} {\bibfield  {journal} {\bibinfo  {journal}
  {Phys. Rev. A}\ }\textbf {\bibinfo {volume} {71}},\ \bibinfo {pages} {055801}
  (\bibinfo {year} {2005})}\BibitemShut {NoStop}%
\bibitem [{\citenamefont {Paraoanu}(2014)}]{Paraoanu:2014}%
  \BibitemOpen
  \bibfield  {author} {\bibinfo {author} {\bibfnamefont {G.~S.}\ \bibnamefont
  {Paraoanu}},\ }\href {\doibase 10.1007/s10909-014-1175-8} {\bibfield
  {journal} {\bibinfo  {journal} {Journal of Low Temperature Physics}\ }\textbf
  {\bibinfo {volume} {175}},\ \bibinfo {pages} {633} (\bibinfo {year}
  {2014})}\BibitemShut {NoStop}%
\bibitem [{\citenamefont {Wendin}(2017)}]{Wendin:2017}%
  \BibitemOpen
  \bibfield  {author} {\bibinfo {author} {\bibfnamefont {G.}~\bibnamefont
  {Wendin}},\ }\href {\doibase 10.1088/1361-6633/aa7e1a} {\bibfield  {journal}
  {\bibinfo  {journal} {Reports on Progress in Physics}\ }\textbf {\bibinfo
  {volume} {80}},\ \bibinfo {pages} {106001} (\bibinfo {year}
  {2017})}\BibitemShut {NoStop}%
\bibitem [{\citenamefont {Krantz}\ \emph {et~al.}(2019)\citenamefont {Krantz},
  \citenamefont {Kjaergaard}, \citenamefont {Yan}, \citenamefont {Orlando},
  \citenamefont {Gustavsson},\ and\ \citenamefont
  {Oliver}}]{Krantz:Kjaergaard:2019}%
  \BibitemOpen
  \bibfield  {author} {\bibinfo {author} {\bibfnamefont {P.}~\bibnamefont
  {Krantz}}, \bibinfo {author} {\bibfnamefont {M.}~\bibnamefont {Kjaergaard}},
  \bibinfo {author} {\bibfnamefont {F.}~\bibnamefont {Yan}}, \bibinfo {author}
  {\bibfnamefont {T.~P.}\ \bibnamefont {Orlando}}, \bibinfo {author}
  {\bibfnamefont {S.}~\bibnamefont {Gustavsson}}, \ and\ \bibinfo {author}
  {\bibfnamefont {W.~D.}\ \bibnamefont {Oliver}},\ }\href {\doibase
  10.1063/1.5089550} {\bibfield  {journal} {\bibinfo  {journal} {Applied
  Physics Reviews}\ }\textbf {\bibinfo {volume} {6}},\ \bibinfo {pages}
  {021318} (\bibinfo {year} {2019})},\ \Eprint
  {http://arxiv.org/abs/https://doi.org/10.1063/1.5089550}
  {https://doi.org/10.1063/1.5089550} \BibitemShut {NoStop}%
\bibitem [{\citenamefont {Tuorila}\ \emph {et~al.}(2010)\citenamefont
  {Tuorila}, \citenamefont {Silveri}, \citenamefont {Sillanp\"a\"a},
  \citenamefont {Thuneberg}, \citenamefont {Makhlin},\ and\ \citenamefont
  {Hakonen}}]{Tuorila:Silveri:2010}%
  \BibitemOpen
  \bibfield  {author} {\bibinfo {author} {\bibfnamefont {J.}~\bibnamefont
  {Tuorila}}, \bibinfo {author} {\bibfnamefont {M.}~\bibnamefont {Silveri}},
  \bibinfo {author} {\bibfnamefont {M.}~\bibnamefont {Sillanp\"a\"a}}, \bibinfo
  {author} {\bibfnamefont {E.}~\bibnamefont {Thuneberg}}, \bibinfo {author}
  {\bibfnamefont {Y.}~\bibnamefont {Makhlin}}, \ and\ \bibinfo {author}
  {\bibfnamefont {P.}~\bibnamefont {Hakonen}},\ }\href {\doibase
  10.1103/PhysRevLett.105.257003} {\bibfield  {journal} {\bibinfo  {journal}
  {Phys. Rev. Lett.}\ }\textbf {\bibinfo {volume} {105}},\ \bibinfo {pages}
  {257003} (\bibinfo {year} {2010})}\BibitemShut {NoStop}%
\bibitem [{\citenamefont {Pietik\"ainen}\ \emph {et~al.}(2017)\citenamefont
  {Pietik\"ainen}, \citenamefont {Danilin}, \citenamefont {Kumar},
  \citenamefont {Veps\"al\"ainen}, \citenamefont {Golubev}, \citenamefont
  {Tuorila},\ and\ \citenamefont {Paraoanu}}]{Pietikainen:Danilin:2017}%
  \BibitemOpen
  \bibfield  {author} {\bibinfo {author} {\bibfnamefont {I.}~\bibnamefont
  {Pietik\"ainen}}, \bibinfo {author} {\bibfnamefont {S.}~\bibnamefont
  {Danilin}}, \bibinfo {author} {\bibfnamefont {K.~S.}\ \bibnamefont {Kumar}},
  \bibinfo {author} {\bibfnamefont {A.}~\bibnamefont {Veps\"al\"ainen}},
  \bibinfo {author} {\bibfnamefont {D.~S.}\ \bibnamefont {Golubev}}, \bibinfo
  {author} {\bibfnamefont {J.}~\bibnamefont {Tuorila}}, \ and\ \bibinfo
  {author} {\bibfnamefont {G.~S.}\ \bibnamefont {Paraoanu}},\ }\href {\doibase
  10.1103/PhysRevB.96.020501} {\bibfield  {journal} {\bibinfo  {journal} {Phys.
  Rev. B}\ }\textbf {\bibinfo {volume} {96}},\ \bibinfo {pages} {020501}
  (\bibinfo {year} {2017})}\BibitemShut {NoStop}%
\bibitem [{\citenamefont {Pietik{\"a}inen}\ \emph {et~al.}(2018)\citenamefont
  {Pietik{\"a}inen}, \citenamefont {Danilin}, \citenamefont {Kumar},
  \citenamefont {Tuorila},\ and\ \citenamefont
  {Paraoanu}}]{Pietikainen:Danilin:2018}%
  \BibitemOpen
  \bibfield  {author} {\bibinfo {author} {\bibfnamefont {I.}~\bibnamefont
  {Pietik{\"a}inen}}, \bibinfo {author} {\bibfnamefont {S.}~\bibnamefont
  {Danilin}}, \bibinfo {author} {\bibfnamefont {K.~S.}\ \bibnamefont {Kumar}},
  \bibinfo {author} {\bibfnamefont {J.}~\bibnamefont {Tuorila}}, \ and\
  \bibinfo {author} {\bibfnamefont {G.~S.}\ \bibnamefont {Paraoanu}},\ }\href
  {\doibase 10.1007/s10909-018-1857-8} {\bibfield  {journal} {\bibinfo
  {journal} {Journal of Low Temperature Physics}\ }\textbf {\bibinfo {volume}
  {191}},\ \bibinfo {pages} {354} (\bibinfo {year} {2018})}\BibitemShut
  {NoStop}%
\bibitem [{\citenamefont {Langford}\ \emph {et~al.}(2017)\citenamefont
  {Langford}, \citenamefont {Sagastizabal}, \citenamefont {Kounalakis},
  \citenamefont {Dickel}, \citenamefont {Bruno}, \citenamefont {Luthi},
  \citenamefont {Thoen}, \citenamefont {Endo},\ and\ \citenamefont
  {DiCarlo}}]{Langford:Sagasrizabal:2017}%
  \BibitemOpen
  \bibfield  {author} {\bibinfo {author} {\bibfnamefont {N.~K.}\ \bibnamefont
  {Langford}}, \bibinfo {author} {\bibfnamefont {R.}~\bibnamefont
  {Sagastizabal}}, \bibinfo {author} {\bibfnamefont {M.}~\bibnamefont
  {Kounalakis}}, \bibinfo {author} {\bibfnamefont {C.}~\bibnamefont {Dickel}},
  \bibinfo {author} {\bibfnamefont {A.}~\bibnamefont {Bruno}}, \bibinfo
  {author} {\bibfnamefont {F.}~\bibnamefont {Luthi}}, \bibinfo {author}
  {\bibfnamefont {D.~J.}\ \bibnamefont {Thoen}}, \bibinfo {author}
  {\bibfnamefont {A.}~\bibnamefont {Endo}}, \ and\ \bibinfo {author}
  {\bibfnamefont {L.}~\bibnamefont {DiCarlo}},\ }\href {\doibase
  10.1038/s41467-017-01061-x} {\bibfield  {journal} {\bibinfo  {journal}
  {Nature Communications}\ }\textbf {\bibinfo {volume} {8}},\ \bibinfo {pages}
  {1715} (\bibinfo {year} {2017})}\BibitemShut {NoStop}%
\bibitem [{\citenamefont {Braum{\"u}ller}\ \emph {et~al.}(2017)\citenamefont
  {Braum{\"u}ller}, \citenamefont {Marthaler}, \citenamefont {Schneider},
  \citenamefont {Stehli}, \citenamefont {Rotzinger}, \citenamefont {Weides},\
  and\ \citenamefont {Ustinov}}]{Braumueller:Marthaler:2017}%
  \BibitemOpen
  \bibfield  {author} {\bibinfo {author} {\bibfnamefont {J.}~\bibnamefont
  {Braum{\"u}ller}}, \bibinfo {author} {\bibfnamefont {M.}~\bibnamefont
  {Marthaler}}, \bibinfo {author} {\bibfnamefont {A.}~\bibnamefont
  {Schneider}}, \bibinfo {author} {\bibfnamefont {A.}~\bibnamefont {Stehli}},
  \bibinfo {author} {\bibfnamefont {H.}~\bibnamefont {Rotzinger}}, \bibinfo
  {author} {\bibfnamefont {M.}~\bibnamefont {Weides}}, \ and\ \bibinfo {author}
  {\bibfnamefont {A.~V.}\ \bibnamefont {Ustinov}},\ }\href {\doibase
  10.1038/s41467-017-00894-w} {\bibfield  {journal} {\bibinfo  {journal}
  {Nature Communications}\ }\textbf {\bibinfo {volume} {8}},\ \bibinfo {pages}
  {779} (\bibinfo {year} {2017})}\BibitemShut {NoStop}%
\bibitem [{\citenamefont {Li}\ \emph {et~al.}(2013)\citenamefont {Li},
  \citenamefont {Silveri}, \citenamefont {Kumar}, \citenamefont {Pirkkalainen},
  \citenamefont {Veps{\"a}l{\"a}inen}, \citenamefont {Chien}, \citenamefont
  {Tuorila}, \citenamefont {Sillanp{\"a}{\"a}}, \citenamefont {Hakonen},
  \citenamefont {Thuneberg},\ and\ \citenamefont {Paraoanu}}]{Li:Silveri:2013}%
  \BibitemOpen
  \bibfield  {author} {\bibinfo {author} {\bibfnamefont {J.}~\bibnamefont
  {Li}}, \bibinfo {author} {\bibfnamefont {M.~P.}\ \bibnamefont {Silveri}},
  \bibinfo {author} {\bibfnamefont {K.~S.}\ \bibnamefont {Kumar}}, \bibinfo
  {author} {\bibfnamefont {J.~M.}\ \bibnamefont {Pirkkalainen}}, \bibinfo
  {author} {\bibfnamefont {A.}~\bibnamefont {Veps{\"a}l{\"a}inen}}, \bibinfo
  {author} {\bibfnamefont {W.~C.}\ \bibnamefont {Chien}}, \bibinfo {author}
  {\bibfnamefont {J.}~\bibnamefont {Tuorila}}, \bibinfo {author} {\bibfnamefont
  {M.~A.}\ \bibnamefont {Sillanp{\"a}{\"a}}}, \bibinfo {author} {\bibfnamefont
  {P.~J.}\ \bibnamefont {Hakonen}}, \bibinfo {author} {\bibfnamefont {E.~V.}\
  \bibnamefont {Thuneberg}}, \ and\ \bibinfo {author} {\bibfnamefont {G.~S.}\
  \bibnamefont {Paraoanu}},\ }\href {\doibase 10.1038/ncomms2383} {\bibfield
  {journal} {\bibinfo  {journal} {Nature Communications}\ }\textbf {\bibinfo
  {volume} {4}},\ \bibinfo {pages} {1420} (\bibinfo {year} {2013})}\BibitemShut
  {NoStop}%
\bibitem [{\citenamefont {Forn-D{\'\i}az}\ \emph {et~al.}(2017)\citenamefont
  {Forn-D{\'\i}az}, \citenamefont {Garc{\'\i}a-Ripoll}, \citenamefont
  {Peropadre}, \citenamefont {Orgiazzi}, \citenamefont {Yurtalan},
  \citenamefont {Belyansky}, \citenamefont {Wilson},\ and\ \citenamefont
  {Lupascu}}]{FronDiaz:Garcia-Ripoll:2017}%
  \BibitemOpen
  \bibfield  {author} {\bibinfo {author} {\bibfnamefont {P.}~\bibnamefont
  {Forn-D{\'\i}az}}, \bibinfo {author} {\bibfnamefont {J.~J.}\ \bibnamefont
  {Garc{\'\i}a-Ripoll}}, \bibinfo {author} {\bibfnamefont {B.}~\bibnamefont
  {Peropadre}}, \bibinfo {author} {\bibfnamefont {J.~L.}\ \bibnamefont
  {Orgiazzi}}, \bibinfo {author} {\bibfnamefont {M.~A.}\ \bibnamefont
  {Yurtalan}}, \bibinfo {author} {\bibfnamefont {R.}~\bibnamefont {Belyansky}},
  \bibinfo {author} {\bibfnamefont {C.~M.}\ \bibnamefont {Wilson}}, \ and\
  \bibinfo {author} {\bibfnamefont {A.}~\bibnamefont {Lupascu}},\ }\href
  {\doibase 10.1038/nphys3905} {\bibfield  {journal} {\bibinfo  {journal}
  {Nature Physics}\ }\textbf {\bibinfo {volume} {13}},\ \bibinfo {pages} {39}
  (\bibinfo {year} {2017})}\BibitemShut {NoStop}%
\bibitem [{\citenamefont {Bosman}\ \emph {et~al.}(2017)\citenamefont {Bosman},
  \citenamefont {Gely}, \citenamefont {Singh}, \citenamefont {Bruno},
  \citenamefont {Bothner},\ and\ \citenamefont {Steele}}]{Bosman:Gely:2017}%
  \BibitemOpen
  \bibfield  {author} {\bibinfo {author} {\bibfnamefont {S.~J.}\ \bibnamefont
  {Bosman}}, \bibinfo {author} {\bibfnamefont {M.~F.}\ \bibnamefont {Gely}},
  \bibinfo {author} {\bibfnamefont {V.}~\bibnamefont {Singh}}, \bibinfo
  {author} {\bibfnamefont {A.}~\bibnamefont {Bruno}}, \bibinfo {author}
  {\bibfnamefont {D.}~\bibnamefont {Bothner}}, \ and\ \bibinfo {author}
  {\bibfnamefont {G.~A.}\ \bibnamefont {Steele}},\ }\href {\doibase
  10.1038/s41534-017-0046-y} {\bibfield  {journal} {\bibinfo  {journal} {npj
  Quantum Information}\ }\textbf {\bibinfo {volume} {3}},\ \bibinfo {pages}
  {46} (\bibinfo {year} {2017})}\BibitemShut {NoStop}%
\bibitem [{\citenamefont {Wallraff}\ \emph {et~al.}(2004)\citenamefont
  {Wallraff}, \citenamefont {Schuster}, \citenamefont {Blais}, \citenamefont
  {Frunzio}, \citenamefont {Huang}, \citenamefont {Majer}, \citenamefont
  {Kumar}, \citenamefont {Girvin},\ and\ \citenamefont
  {Schoelkopf}}]{Wallraff:Schuster:2004}%
  \BibitemOpen
  \bibfield  {author} {\bibinfo {author} {\bibfnamefont {A.}~\bibnamefont
  {Wallraff}}, \bibinfo {author} {\bibfnamefont {D.~I.}\ \bibnamefont
  {Schuster}}, \bibinfo {author} {\bibfnamefont {A.}~\bibnamefont {Blais}},
  \bibinfo {author} {\bibfnamefont {L.}~\bibnamefont {Frunzio}}, \bibinfo
  {author} {\bibfnamefont {R.~S.}\ \bibnamefont {Huang}}, \bibinfo {author}
  {\bibfnamefont {J.}~\bibnamefont {Majer}}, \bibinfo {author} {\bibfnamefont
  {S.}~\bibnamefont {Kumar}}, \bibinfo {author} {\bibfnamefont {S.~M.}\
  \bibnamefont {Girvin}}, \ and\ \bibinfo {author} {\bibfnamefont {R.~J.}\
  \bibnamefont {Schoelkopf}},\ }\href {\doibase 10.1038/nature02851} {\bibfield
   {journal} {\bibinfo  {journal} {Nature}\ }\textbf {\bibinfo {volume}
  {431}},\ \bibinfo {pages} {162} (\bibinfo {year} {2004})}\BibitemShut
  {NoStop}%
\end{thebibliography}%


\appendix

\onecolumngrid

\newpage

\section{Time evolution in the ultrastrong coupling regime through the symplectic formalism}\label{time:evolution:strong:coupling:regime}
The (two-mode) bosonic Hamiltonian $\hat{H}$ that us quadratic in the creation and annihilation operators, or equivalently in the quadratures, has matrix representation $\boldsymbol{H}$ that is obtained through the definition $\hat{H}=\frac{\hbar}{2}\,\mathbb{X}^\dag\,\boldsymbol{H}\,\mathbb{X}$. The generic expression for $\boldsymbol{H}$ is
\begin{align}
\boldsymbol{H}
=
\begin{pmatrix}
\boldsymbol{U} & \boldsymbol{V} \\
\boldsymbol{V}^* & \boldsymbol{U}^*
\end{pmatrix}.
\end{align}
Here, $\boldsymbol{U}=\boldsymbol{U}^\dag$ and $\boldsymbol{V}=\boldsymbol{V}^{\textrm{Tp}}$.

The Hamiltonian induces time evolution in the symplectic formalism through the symplectic matrix
\begin{align}\label{time:evolution:matrix:strong:coupling:appendix}
\boldsymbol{S}(t):=\exp[\boldsymbol{\Omega}\,\boldsymbol{H}\,t],
\end{align}
where we have defined the symplectic form $\boldsymbol{\Omega}:=-i\,\textrm{diag}(1,1,-1,-1)$ and we recall that symplectic matrices $\boldsymbol{S}$ satisfy the property $\boldsymbol{S}\,\boldsymbol{\Omega}\,\boldsymbol{S}^\dag=\boldsymbol{S}^\dag\,\boldsymbol{\Omega}\,\boldsymbol{S}=\boldsymbol{\Omega}$.

Williamson's theorem guarantees that a positive definite, Hermitian or symmetric $2N\times2N$ matrix $\boldsymbol{H}$, such as the Hamitlonian, can be put in diagonal form through the relation $\boldsymbol{H}=\boldsymbol{s}^\dag\,\tilde{\boldsymbol{\kappa}}\,\boldsymbol{s}$, where $\boldsymbol{s}$ is  a symplectic matrix \cite{Williamson:1923}. Here, we have that $\tilde{\boldsymbol{\kappa}}$ is the symplectic form of $\boldsymbol{H}$.\footnote{In the literature, the symplectic form \textit{of the covariance matrix} is usually denoted by $\boldsymbol{\nu}_\oplus$. } Therefore, we have 
\begin{align}\label{main:expression:symplectic:matrix}
\boldsymbol{S}(t)&=\exp[\boldsymbol{\Omega}\,\boldsymbol{H}\,t]\nonumber\\
&=\exp[\boldsymbol{\Omega}\,\boldsymbol{s}^\dag\,\tilde{\boldsymbol{\kappa}}\,\boldsymbol{s}\,t]\nonumber\\
&=\exp[\boldsymbol{s}^{-1}\,\boldsymbol{s}\,\boldsymbol{\Omega}\,\boldsymbol{s}^\dag\,\tilde{\boldsymbol{\kappa}}\,\boldsymbol{s}\,t]\nonumber\\
&=\exp[\boldsymbol{s}^{-1}\,\boldsymbol{\Omega}\,\tilde{\boldsymbol{\kappa}}\,\boldsymbol{s}\,t]\nonumber\\
&=\boldsymbol{s}^{-1}\,\exp[\boldsymbol{\Omega}\,\tilde{\boldsymbol{\kappa}}\,t]\,\boldsymbol{s}\nonumber\\
&=-\boldsymbol{\Omega}\,\boldsymbol{s}^\dag\,\boldsymbol{\Omega}\,\exp[\boldsymbol{\Omega}\,\tilde{\boldsymbol{\kappa}}\,t]\,\boldsymbol{s},
\end{align}
where we have used the property $\boldsymbol{s}^{-1}=-\boldsymbol{\Omega}\,\boldsymbol{s}^\dag\,\boldsymbol{\Omega}$ derived from the definition $\boldsymbol{s}\,\boldsymbol{\Omega}\,\boldsymbol{s}^\dag=\boldsymbol{\Omega}$ of symplectic matrices. We emphasize that the symplectic matrix $\boldsymbol{s}$ is the one that diagonalises the Hamitlonian matrix $\boldsymbol{H}$. We also have that $\tilde{\boldsymbol{\kappa}}=\text{diag}(\kappa_+,\kappa_-,\kappa_+,\kappa_-)$, and these elements are known as the symplectic eigenvalues \cite{Adesso:Ragy:2014} of the matrix $\boldsymbol{H}$.

Therefore, we have found that
\begin{align}\label{time:evolution:matrix:strong:coupling:final:form:appendix}
\boldsymbol{S}(t)&=-\boldsymbol{\Omega}\,\boldsymbol{s}^\dag\,\boldsymbol{\Omega}\,\exp[\boldsymbol{\Omega}\,\tilde{\boldsymbol{\kappa}}\,t]\,\boldsymbol{s}.
\end{align}
Notice that the expression \eqref{time:evolution:matrix:strong:coupling:final:form:appendix} holds for \textit{any} time independent and quadratic Hamiltonian.
 
Defining the symplectic matrices $\boldsymbol{S}(t)$ as
\begin{align}\label{symplectic:matrices:appendix}
\boldsymbol{S}(t)
=
\begin{pmatrix}
\boldsymbol{A}(t) & \boldsymbol{B}(t) \\
\boldsymbol{B}^*(t) & \boldsymbol{A}^*(t) 
\end{pmatrix},
\quad\quad
\boldsymbol{s}
=
\begin{pmatrix}
\boldsymbol{\alpha} & \boldsymbol{\beta} \\
\boldsymbol{\beta} & \boldsymbol{\alpha} 
\end{pmatrix}
\end{align}
we then obtain
\begin{align}\label{main:transformation:equation}
\boldsymbol{A}(t)&=\boldsymbol{\alpha}^\text{Tp}\,e^{-i\,\boldsymbol{\kappa}\,t}\,\boldsymbol{\alpha}-\boldsymbol{\beta}^\text{Tp}\,e^{i\,\boldsymbol{\kappa}\,t}\,\boldsymbol{\beta}\nonumber\\
\boldsymbol{B}(t)&=\boldsymbol{\alpha}^\text{Tp}\,e^{-i\,\boldsymbol{\kappa}\,t}\,\boldsymbol{\beta}-\boldsymbol{\beta}^\text{Tp}\,e^{i\,\boldsymbol{\kappa}\,t}\,\boldsymbol{\alpha},
\end{align}
where we have defined $\boldsymbol{\kappa}=\text{diag}(\kappa_+,\kappa_-)$ and that, therefore, $\tilde{\boldsymbol{\kappa}}=\boldsymbol{\kappa}\oplus\boldsymbol{\kappa}$. These coefficients have the same functional form of those that appear in the results found in the literature for the resonant case \cite{Estes:Keil:1968}. Needless to say, they also reduce to such coefficients when $\omega_\text{a}=\omega_\text{b}$. Furthermore, notice that $\boldsymbol{A}^{\text{Tp}}(t)=\boldsymbol{A}(t)$ and $\boldsymbol{B}^\dag(t)=-\boldsymbol{B}(t)$.

Together with the main equations \eqref{main:transformation:equation} we need the constraints
\begin{align}\label{main:constraints:equation}
\boldsymbol{U}&=\boldsymbol{\alpha}^\text{Tp}\,\boldsymbol{\kappa}\,\boldsymbol{\alpha}+\boldsymbol{\beta}^\text{Tp}\,\boldsymbol{\kappa}\,\boldsymbol{\beta}\nonumber\\
\boldsymbol{V}&=\boldsymbol{\alpha}^\text{Tp}\,\boldsymbol{\kappa}\,\boldsymbol{\beta}+\boldsymbol{\beta}^\text{Tp}\,\boldsymbol{\kappa}\,\boldsymbol{\alpha},
\end{align}
which are nothing more than the statement that the Hamiltonian $\boldsymbol{H}$ is diagonalised by the symplectic matrix $\boldsymbol{s}$.

Notice that we can multiply the first line of \eqref{main:constraints:equation} on the right by $\boldsymbol{\alpha}^\text{Tp}$ and the second line by $\boldsymbol{\beta}^\text{Tp}$, to obtain
\begin{align}
\boldsymbol{U}\,\boldsymbol{\alpha}^\text{Tp}&=\boldsymbol{\alpha}^\text{Tp}\,\boldsymbol{\kappa}\,\boldsymbol{\alpha}\,\boldsymbol{\alpha}^\text{Tp}+\boldsymbol{\beta}^\text{Tp}\,\boldsymbol{\kappa}\,\boldsymbol{\beta}\,\boldsymbol{\alpha}^\text{Tp}\nonumber\\
\boldsymbol{V}\,\boldsymbol{\beta}^\text{Tp}&=\boldsymbol{\alpha}^\text{Tp}\,\boldsymbol{\kappa}\,\boldsymbol{\beta}\,\boldsymbol{\beta}^\text{Tp}+\boldsymbol{\beta}^\text{Tp}\,\boldsymbol{\kappa}\,\boldsymbol{\alpha}\,\boldsymbol{\beta}^\text{Tp}.
\end{align}
Subtracting and using the Bogoliubov identities we have
\begin{align}
\boldsymbol{U}\,\boldsymbol{\alpha}^\text{Tp}-\boldsymbol{V}\,\boldsymbol{\beta}^\text{Tp}&=\boldsymbol{\alpha}^\text{Tp}\,\boldsymbol{\kappa}.
\end{align}
Inverting the products and repeating we get
\begin{align}
\boldsymbol{U}\,\boldsymbol{\beta}^\text{Tp}-\boldsymbol{V}\,\boldsymbol{\alpha}^\text{Tp}&=-\boldsymbol{\beta}^\text{Tp}\,\boldsymbol{\kappa}.
\end{align}
Taking the transpose we obtain the equivalent sets of constraints
\begin{align}\label{main:constraints:equation:alternative}
\boldsymbol{\alpha}\,\boldsymbol{U}-\boldsymbol{\beta}\,\boldsymbol{V}&=\boldsymbol{\kappa}\,\boldsymbol{\alpha}\nonumber\\
\boldsymbol{\beta}\,\boldsymbol{U}-\boldsymbol{\alpha}\,\boldsymbol{V}&=-\boldsymbol{\kappa}\,\boldsymbol{\beta}.
\end{align}
Finally, we see that the symplectic frequencies $\kappa_\pm$ read
\begin{align}\label{symplectic:frequencies:appendix}
\kappa_\pm^2:=&\frac{1}{2}\left[\omega_\textrm{a}^2+\omega_\textrm{b}^2+2(g_\textrm{bs}^2-g_\textrm{sq}^2)-\lambda_\textrm{a}^2-\lambda_\textrm{b}^2\pm\left(\left(\omega_\textrm{a}^2+\omega_\textrm{b}^2+2 (g_\textrm{bs}^2-g_\textrm{sq}^2)-\lambda_\textrm{a}^2-\lambda_\textrm{b}^2\right)^2-4 \omega_\textrm{a}^2 \omega_\textrm{b}^2+4 \omega_\textrm{a}^2 \lambda_\textrm{b}^2+4\omega_\textrm{b}^2 \lambda_\textrm{a}^2\right.\right.\nonumber\\
&\left.\left.+8\omega_\textrm{a} \omega_\textrm{b} g_\textrm{bs}^2+8 \omega_\textrm{a} \omega_\textrm{b}\,g_\textrm{sq}^2-16\,(\omega_\textrm{a}\,\lambda_\textrm{b}+\omega_\textrm{b}\,\lambda_\textrm{a})\,g_\textrm{bs}\,g_\textrm{sq}+8(g_\textrm{bs}^2+g_\textrm{sq}^2)\,\lambda_\textrm{a}\,\lambda_\textrm{b}-4\,(g_\textrm{bs}^2-g_\textrm{sq}^2)^2-4\lambda_\textrm{a}^2\,\lambda_\textrm{b}^2\right)^{1/2}\right]
\end{align}
in the general case. When $,\lambda_\textrm{a}=\lambda_\textrm{b}=0$ they reduce to
\begin{align}\label{normal:mode:frequencies:strong:coupling:general:appendix}
\kappa_\pm^2=&\frac{1}{2}\left[\omega_\text{a}^2+\omega_\text{b}^2+2\,(g_\text{bs}^2-g_\text{sq}^2)\pm\left((\omega_\text{a}^2-\omega_\text{b}^2)^2+8\,\omega_\text{a}\,\omega_\text{b}\,(g_\text{bs}^2+g_\text{sq}^2)+4(\omega_\text{a}^2+\omega_\text{b}^2)\,(g_\text{bs}^2-g_\text{sq}^2)\right)^{\frac{1}{2}}\right].
\end{align}

\section{Time evolution through the symplectic formalism: $g_\text{sq}=g_\text{bs}=g$}\label{time:evolution:strong:coupling:regime:real}
Let us assume here that $g_\text{sq}=g_\text{bs}=g$. In this case we can find a solution for \eqref{main:constraints:equation} using \eqref{main:constraints:equation:alternative}.
First of all we write the Bogoliubov matrices as
\begin{align}\label{bogoliubov:matrix:appendix}
\boldsymbol{\alpha}
=
\begin{pmatrix}
\alpha_{11} & \alpha_{12} \\
\alpha_{21} & \alpha_{22}
\end{pmatrix},
\quad\quad
\boldsymbol{\beta}
=
\begin{pmatrix}
\beta_{11} & \beta_{12} \\
\beta_{21} & \beta_{22}
\end{pmatrix}.
\end{align}
Lengthy algebra allows us first to show that
\begin{align}\label{bogoliubov:coefficients:appendix}
\alpha_{11}&=\frac{\kappa_++\omega_\text{a}-\lambda_\textrm{a}}{2 \sqrt{\kappa_+\,(\omega_\text{a}-\lambda_\textrm{a})}}\cos\theta & \beta_{11}&=-\frac{\omega_\text{a}-\kappa_+-\lambda_\textrm{a}}{2 \sqrt{\kappa_+\,(\omega_\text{a}-\lambda_\textrm{a})}}\cos\theta \nonumber\\
 \alpha_{12}&=\frac{\kappa_++\omega_\text{b}-\lambda_\textrm{b}}{2 \sqrt{\kappa_+\,(\omega_\text{b}-\lambda_\textrm{b})}}\sin\theta & \beta_{12}&=-\frac{\omega_\text{b}-\kappa_+-\lambda_\textrm{b}}{2 \sqrt{\kappa_+\,(\omega_\text{b}-\lambda_\textrm{b})}}\sin\theta \nonumber\\
 \alpha_{21}&=\frac{\kappa_-+\omega_\text{a}-\lambda_\textrm{a}}{2 \sqrt{\kappa_-\,(\omega_\text{a}-\lambda_\textrm{a})}}\sin\theta & \beta_{21}&=-\frac{\omega_\text{a}-\kappa_--\lambda_\textrm{a}}{2\sqrt{\kappa_-\,(\omega_\text{a}-\lambda_\textrm{a})}}\sin\theta \nonumber\\
 \alpha_{22}&=-\frac{\kappa_-+\omega_\text{b}-\lambda_\textrm{b}}{2\sqrt{\kappa_-\,(\omega_\text{b}-\lambda_\textrm{b})}}\cos\theta & \beta_{22}&=-\frac{\omega_\text{b}-\kappa_--\lambda_\textrm{b}}{2\sqrt{\kappa_-\,(\omega_\text{b}-\lambda_\textrm{b})}}\cos\theta,
\end{align}
which we supplement with the definition $\theta$, which reads
\begin{align}
\tan(2\,\theta):=&\frac{\sqrt{(\sqrt{\left(\omega_\textrm{a}^2-\omega_\textrm{b}^2+\lambda_\textrm{b}^2-\lambda_\textrm{a}^2\right)^2+16\,\left(\omega_\textrm{a}\,\omega_\textrm{b}+\lambda_\textrm{a}\,\lambda_\textrm{b}\right)\,g^2}-(\lambda_\textrm{b}^2-\lambda_\textrm{a}^2))^2-(\omega_\textrm{a}^2-\omega_\textrm{b}^2)^2}}{(\omega_\text{a}^2-\omega_\textrm{b}^2)}.
\end{align}
Notice that of $0\leq2\theta\leq\pi$ and $\alpha_{11}^2-\beta_{11}^2=\alpha_{22}^2-\beta_{22}^2=\cos^2\theta$ and $\alpha_{12}^2-\beta_{12}^2=\alpha_{21}^2-\beta_{21}^2=\sin^2\theta$.

Finally, focussing on $\lambda_\textrm{a}=\lambda_\textrm{b}=0$ for simplicity of presentation, this allows us to obtain
\begin{align}\label{coefficients:total}
A_{11}(t)=&\left[\cos^2\theta\,\cos(\kappa_+\,t)+\sin^2\theta\,\cos(\kappa_-\,t)\right]-\frac{i}{2}\left[\frac{\kappa_+^2+\omega_\text{a}^2}{\kappa_+\,\omega_\text{a}}\cos^2\theta\,\sin(\kappa_+\,t)+\frac{\kappa_-^2+\omega_\text{a}^2}{\kappa_-\,\omega_\text{a}}\sin^2\theta\,\sin(\kappa_-\,t)\right]\nonumber\\
A_{12}(t)=&\frac{\omega_\text{a}+\omega_\textrm{b}}{4\,\sqrt{\omega_\textrm{b}\,\omega_\text{a}}}\,\sin(2\,\theta)\,\left[\cos(\kappa_+\,t)-\cos(\kappa_-\,t)\right]-i\,\frac{\sin(2\,\theta)}{4\,\sqrt{\omega_\textrm{b}\,\omega_\text{a}}}\left[\frac{\kappa_+^2+\omega_\textrm{b}\,\omega_\text{a}}{\kappa_+}\,\sin(\kappa_+\,t)-\frac{\kappa_-^2+\omega_\textrm{b}\,\omega_\text{a}}{\kappa_-}\,\sin(\kappa_-\,t)\right]\nonumber\\
A_{21}(t)=&A_{12}(t)\nonumber\\
A_{22}(t)=&\left[\sin^2\theta\,\cos(\kappa_+\,t)+\cos^2\theta\,\cos(\kappa_-\,t)\right]-\frac{i}{2}\left[\frac{\kappa_+^2+\omega_\textrm{b}^2}{\kappa_+\,\omega_\textrm{b}}\sin^2\theta\,\sin(\kappa_+\,t)+\frac{\kappa_-^2+\omega_\textrm{b}^2}{\kappa_-\,\omega_\textrm{b}}\cos^2\theta\,\sin(\kappa_-\,t)\right]\nonumber\\
B_{11}(t)=&-\frac{i}{2}\left[\frac{\kappa_+^2-\omega_\text{a}^2}{\kappa_+\,\omega_\text{a}}\cos^2\theta\,\sin(\kappa_+\,t)+\frac{\kappa_-^2-\omega_\text{a}^2}{\kappa_-\,\omega_\text{a}}\sin^2\theta\,\sin(\kappa_-\,t)\right]\nonumber\\
B_{12}(t)=&\frac{\omega_\text{a}-\omega_\textrm{b}}{4\,\sqrt{\omega_\textrm{b}\,\omega_\text{a}}}\,\sin(2\,\theta)\,\left[\cos(\kappa_+\,t)-\cos(\kappa_-\,t)\right]-i\,\frac{\sin(2\,\theta)}{4\,\sqrt{\omega_\textrm{b}\,\omega_\text{a}}}\left[\frac{\kappa_+^2-\omega_\textrm{b}\,\omega_\text{a}}{\kappa_+}\,\sin(\kappa_+\,t)-\frac{\kappa_-^2-\omega_\textrm{b}\,\omega_\text{a}}{\kappa_-}\,\sin(\kappa_-\,t)\right]\nonumber\\
B_{21}(t)=&-B^*_{12}(t)\nonumber\\
B_{22}(t)=&-\frac{i}{2}\left[\frac{\kappa_+^2-\omega_\textrm{b}^2}{\kappa_+\,\omega}\sin^2\theta\,\sin(\kappa_+\,t)+\frac{\kappa_-^2-\omega_\textrm{b}^2}{\kappa_-\,\omega_\textrm{b}}\cos^2\theta\,\sin(\kappa_-\,t)\right].
\end{align}

\section{Time evolution channel decomposition}\label{channel:decomposition:appendix}
In this section we wish to employ a result that shows that any linear optical transformation of $N$ bosonic modes, i.e., any unitary induced by a quadratic Hamiltonian, can be decomposed into a product of a generalized mode mixing, a single mode squeezer acting on all modes, followed by another generalized mode mixer \cite{Braunstein:2005}.

In particular, we start from the transformation \eqref{main:expression:symplectic:matrix} and its form
\begin{align}
\boldsymbol{S}(t)&=\boldsymbol{s}^{-1}\,\exp[\boldsymbol{\Omega}\,\tilde{\boldsymbol{\kappa}}\,t]\,\boldsymbol{s},
\end{align}
and make the ansatz
\begin{align}
\boldsymbol{s}=\boldsymbol{o}(\varphi) \boldsymbol{s}_\text{q}(\underline{r}) \boldsymbol{o}(\phi),
\end{align}
where $\boldsymbol{o}(\psi)$ is an orthogonal matrix and $\boldsymbol{s}_\text{q}(\underline{r})$ is the squeezing matrix \cite{Braunstein:2005}. All matrices have real entries and have the explicit form
\begin{align}\label{decomposition:explicit:expression:appendix}
\boldsymbol{o}(\psi)
:=
\begin{pmatrix}
\cos\psi & \sin\psi & 0 & 0\\
-\sin\psi & \cos\psi & 0 & 0\\
0 & 0 & \cos\psi & \sin\psi\\
0 & 0 & -\sin\psi & \cos\psi
\end{pmatrix},
\quad\quad
\boldsymbol{s}_\text{q}(\underline{r})
:=
\begin{pmatrix}
\cosh r_\text{a} & 0 & \sinh r_\text{a} & 0\\
0 & \cosh r_\text{b} & 0 & \sinh r_\text{b}\\
\sinh r_\text{a} & 0 & \cosh r_\text{a} & 0\\
0 & \sinh r_\text{b} &0 & \cosh r_\text{b}
\end{pmatrix}.
\end{align}
Note the following algebraic properties of these matrices: $\boldsymbol{o}(\psi)\,\boldsymbol{o}^\text{Tp}(\psi)=\mathds{1}$, $\boldsymbol{o}^{-1}(\psi)=\boldsymbol{o}(-\psi)$, $\boldsymbol{s}_\text{q}(\underline{r})=\boldsymbol{s}_\text{q}^\text{Tp}(\underline{r})$ and $\boldsymbol{s}^{-1}_\text{q}(\underline{r})=\boldsymbol{s}_\text{q}(-\underline{r})$.
This implies that we have 
\begin{align}
\boldsymbol{S}(t)&=\boldsymbol{o}(-\phi) \boldsymbol{s}_\text{q}(-\underline{r}) \boldsymbol{o}(-\varphi)\,\exp[\boldsymbol{\Omega}\,\tilde{\boldsymbol{\kappa}}\,t]\,\boldsymbol{o}(\varphi) \boldsymbol{s}_\text{q}(\underline{r}) \boldsymbol{o}(\phi),
\end{align}
which is our main decomposition result.

We then need to match the decomposition $\boldsymbol{s}=\boldsymbol{o}(\varphi) \boldsymbol{s}_\text{q}(\underline{r}) \boldsymbol{o}(\phi)$ to the explicit form found in \eqref{symplectic:matrices:appendix}, with coefficients \eqref{bogoliubov:matrix:appendix} that read \eqref{bogoliubov:coefficients:appendix}. Lengthy algebra leads us to obtain the expressions
\begin{align}\label{parameter:relations:appendix}
\tan(2\,\phi):=&2\frac{\alpha_{11}\,\alpha_{12}+\alpha_{21}\,\alpha_{22}}{\alpha_{11}^2-\alpha_{12}^2+\alpha_{21}^2-\alpha_{22}^2}\nonumber\\
\tan(2\,\varphi):=&2\frac{\alpha_{11}\,\alpha_{21}+\alpha_{12}\,\alpha_{22}}{\alpha_{11}^2+\alpha_{12}^2-\alpha_{21}^2-\alpha_{22}^2}\nonumber\\
\cosh^2(r_\text{a}):=&\frac{1}{2}\left[\alpha_{11}^2+\alpha_{12}^2+\alpha_{21}^2+\alpha_{22}^2+\sqrt{\left(\alpha_{11}^2-\alpha_{12}^2+\alpha_{21}^2-\alpha_{22}^2\right)^2+4\,\left(\alpha_{11}\alpha_{12}+\alpha_{21}\alpha_{22}\right)^2}\right]\nonumber\\
\cosh^2(r_\text{b}):=&\frac{1}{2}\left[\alpha_{11}^2+\alpha_{12}^2+\alpha_{21}^2+\alpha_{22}^2-\sqrt{\left(\alpha_{11}^2-\alpha_{12}^2+\alpha_{21}^2-\alpha_{22}^2\right)^2+4\,\left(\alpha_{11}\alpha_{12}+\alpha_{21}\alpha_{22}\right)^2}\right],
\end{align}
which prove that the ansatz provides a solution to the decomposition problem.

A more illuminating form of \eqref{parameter:relations:appendix}, which highlights the role of the squeezing, can be obtained by manipulating the expressions and using the Bogoliubov identities. We find
\begin{align}\label{parameter:relations:better:appendix}
\tan(2\,\phi):=&2\frac{\alpha_{11}\,\alpha_{12}+\alpha_{21}\,\alpha_{22}}{2\alpha_{21}^2-2\alpha_{12}^2+\beta_{11}^2+\beta_{12}^2-\beta_{21}^2-\beta_{22}^2}\nonumber\\
\tan(2\,\varphi):=&2\frac{\beta_{11}\,\beta_{21}+\beta_{12}\,\beta_{22}}{\beta_{11}^2+\beta_{12}^2-\beta_{21}^2-\beta_{22}^2}\nonumber\\
\sinh^2(r_\text{a}):=&\frac{1}{2}\left[\beta_{11}^2+\beta_{12}^2+\beta_{21}^2+\beta_{22}^2+\sqrt{(\beta_{11}^2-\beta_{22}^2)^2+(\beta_{12}^2-\beta_{21}^2)^2+2(\beta_{11}^2+\beta_{22}^2)(\beta_{12}^2+\beta_{21}^2)+4\beta_{11}\beta_{12}\beta_{21}\beta_{22}}\right]\nonumber\\
\sinh^2(r_\text{b}):=&\frac{1}{2}\left[\beta_{11}^2+\beta_{12}^2+\beta_{21}^2+\beta_{22}^2-\sqrt{(\beta_{11}^2-\beta_{22}^2)^2+(\beta_{12}^2-\beta_{21}^2)^2+2(\beta_{11}^2+\beta_{22}^2)(\beta_{12}^2+\beta_{21}^2)+4\beta_{11}\beta_{12}\beta_{21}\beta_{22}}\right].
\end{align}
It is clear that, when the $\beta_{nm}$ coefficients vanish, there is no squeezing in the overall evolution -- signalled by the fact that, in this case, $r_\text{a}=r_\text{b}=0$.

\section{Superconducting circuits as a platform for quadratic bosonic Hamiltonians}\label{superconducting:circuits:appendix}
\begin{figure}[h!]
    \includegraphics[scale=0.2]{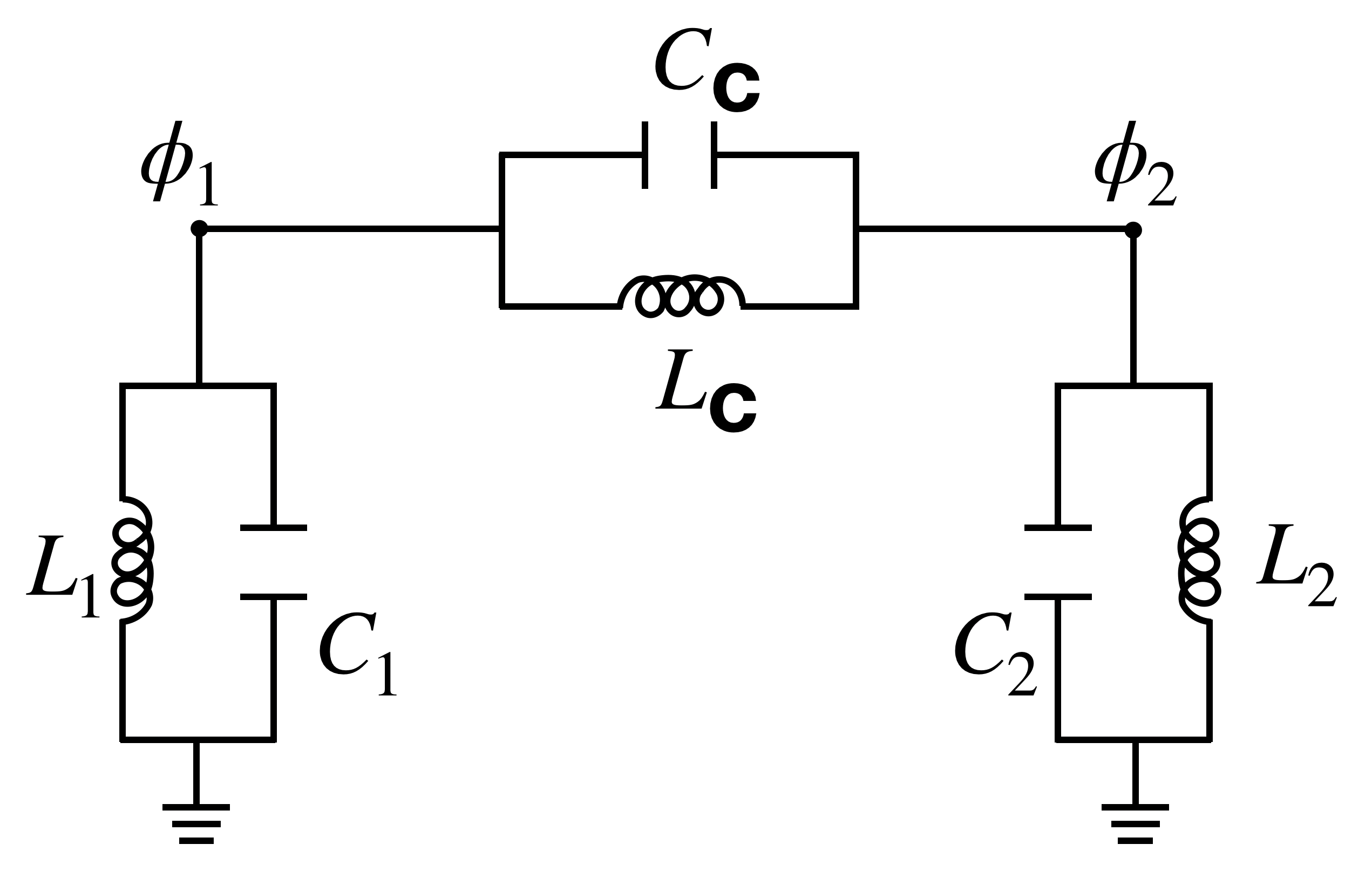}
    \caption{\label{figure:three} Three coupled microwave $LC$ resonators, where $L$ is the inductivity and $C$ the capacity.}
\end{figure}

We proceed here to derive the interaction Hamiltonians presented in \eqref{circuit:couplings} from a basic circuit QED implementation. A pictorial representation of the system considered can be found in \autoref{figure:three}.

We start by considering three LC resonators coupled as depicted. Here, $L$ stands for impedance and the $C$ for capacitance. The variables $\phi_1$ and $\phi_2$ are the fluxes. 

The Lagrangian $\mathcal{L}=\mathcal{L}(\phi_1,\phi_2,\dot{\phi}_1,\dot{\phi}_2)$ of the system reads
\begin{align}\label{lagrangian:appendix}
\mathcal{L}=\frac{1}{2}\,C_1\,\dot{\phi}_1^2-\frac{1}{2}\,\frac{\phi_1^2}{L_1}+\frac{1}{2}\,C_2\,\dot{\phi}_2^2-\frac{1}{2}\,\frac{\phi_2^2}{L_2}+\frac{1}{2}\,C_\textbf{c}\,(\dot{\phi}_1-\dot{\phi}_2)^2-\frac{1}{2}\,\frac{(\phi_1-\phi_2)^2}{L_\textbf{c}}
\end{align}
We can compute the conjugate momenta $\pi_1$ and $\pi_2$ of the variables $\phi_1$ and $\phi_2$ and we find
\begin{align}
\pi_1:=&\frac{\delta\mathcal{L}}{\delta\phi_1}=(C_1+C_\textbf{c})\,\dot{\phi}_1-C_\textbf{c}\,\dot{\phi}_2\nonumber\\
\pi_2:=&\frac{\delta\mathcal{L}}{\delta\phi_2}=-C_\textbf{c}\,\dot{\phi}_1+(C_2+C_\textbf{c})\,\dot{\phi}_2.
\end{align}
In other words, we have 
\begin{align}
\begin{pmatrix}
\pi_1\\
\pi_2
\end{pmatrix}
=
\begin{pmatrix}
C_1+C_\textbf{c} & -C_\textbf{c}\\
-C_\textbf{c} & C_2+C_\textbf{c}
\end{pmatrix}
\begin{pmatrix}
\dot{\phi}_1\\
\dot{\phi}_2
\end{pmatrix},
\end{align}
which implies that
\begin{align}
\begin{pmatrix}
\dot{\phi}_1\\
\dot{\phi}_2
\end{pmatrix}
=
\begin{pmatrix}
\frac{C_2+C_\textbf{c}}{(C_1+C_\textbf{c})\,(C_2+C_\textbf{c})-C_\textbf{c}^2} & \frac{C_\textbf{c}}{(C_1+C_\textbf{c})\,(C_2+C_\textbf{c})-C_\textbf{c}^2}\\
\frac{C_\textbf{c}}{(C_1+C_\textbf{c})\,(C_2+C_\textbf{c})-C_\textbf{c}^2} & \frac{C_1+C_\textbf{c}}{(C_1+C_\textbf{c})\,(C_2+C_\textbf{c})-C_\textbf{c}^2}
\end{pmatrix}
\begin{pmatrix}
\pi_1\\
\pi_2
\end{pmatrix}.
\end{align}
The Hamiltonian $\mathcal{H}=\mathcal{H}(\phi_1,\phi_2,\pi_1,\pi_2)$ is defined as $\mathcal{H}:=\pi_1\,\dot{\phi}_1+\pi_2\,\dot{\phi}_2-\mathcal{L}$.

Lengthy algebra allows us to find
\begin{align}
\mathcal{H}=\frac{\pi_1^2}{2\,\tilde{C}_1}+\frac{\phi_1^2}{2\,\tilde{L}_1}+\frac{\pi_2^2}{2\,\tilde{C}_2}+\frac{\phi_2^2}{2\,\tilde{L}_2}+\tilde{C}_\textbf{c}\,\pi_1\,\pi_2+\frac{1}{\tilde{L}_\text{c}}\,\phi_1\,\phi_2,
\end{align}
where we have defined
\begin{align}
\tilde{C}_1:=&\frac{C_1\,C_2+C_\textbf{c}\,(C_1+C_2)}{C_2+C_\textbf{c}}, & \tilde{C}_2:=&\frac{C_1\,C_2+C_\textbf{c}\,(C_1+C_2)}{C_1+C_\textbf{c}}, & \tilde{C}_\textbf{c}:=&\frac{C_1\,C_2+C_\textbf{c}\,(C_1+C_2)}{C_\textbf{c}}\nonumber\\
\frac{1}{\tilde{L}_1}:=&\frac{1}{L_1}+\frac{1}{L_\textbf{c}}  & \frac{1}{\tilde{L}_2}:=&\frac{1}{L_2}+\frac{1}{L_\textbf{c}} & \tilde{L}_\textbf{c}:=&L_\textbf{c}.
\end{align}
for convenience of presentation.

We now introduce the convenient quantities $\omega_1$, $\omega_2$, $Z_1$, and $Z_2$ defined as 
\begin{align}
\omega_1:=&\frac{1}{\sqrt{\tilde{L}_1\,\tilde{C}_1}}, & Z_1:=&\sqrt{\frac{\tilde{L}_1}{\tilde{C}_1}}, & \phi_{1\text{ZPF}}:=&\sqrt{\frac{Z_1}{2}}, & p_{1\text{ZPF}}:=&\sqrt{\frac{1}{2\,Z_1}}\nonumber\\
\omega_2:=&\frac{1}{\sqrt{\tilde{L}_2\,\tilde{C}_2}}, & Z_2:=&\sqrt{\frac{\tilde{L}_2}{\tilde{C}_2}}, & \phi_{2\text{ZPF}}:=&\sqrt{\frac{Z_2}{2}}, & p_{2\text{ZPF}}:=&\sqrt{\frac{1}{2\,Z_2}},
\end{align}
and perform the following quantization
\begin{align}
\phi_1\rightarrow\hat{\phi}_1=&\phi_{1\text{ZPF}}\,(\hat{a}_1+\hat{a}_1^\dag), & p_1\rightarrow\hat{p}_1=&\frac{1}{i}\,p_{1\text{ZPF}}\,(\hat{a}_2-\hat{a}_2^\dag)\nonumber\\
\phi_2\rightarrow\hat{\phi}_2=&\phi_{1\text{ZPF}}\,(\hat{a}_2+\hat{a}_2^\dag)  & p_2\rightarrow\hat{p}_2=&\frac{1}{i}\,p_{2\text{ZPF}}\,(\hat{a}_2-\hat{a}_2^\dag),
\end{align}
where imposing the canonical commutation relations $[\hat{\phi}_1,\hat{p}_1]=[\hat{\phi}_2,\hat{p}_2]=i$ implies that $[\hat{a}_1,\hat{a}_1^\dag]=[\hat{a}_2,\hat{a}_2^\dag]=1$.

Therefore, we finally obtain the quantized Hamiltonian $\hat{\mathcal{H}}$, which reads
\begin{align}\label{hamiltonioan:CQED:appendix}
\hat{\mathcal{H}}=\omega_1\,\left(\hat{a}_1^\dag\,\hat{a}_1+\frac{1}{2}\right)+\omega_2\,\left(\hat{a}_2^\dag\,\hat{a}_2+\frac{1}{2}\right)-g_\text{C}\,(\hat{a}_1^\dag-\hat{a}_1)\,(\hat{a}_2^\dag-\hat{a}_2)+g_\text{L}\,(\hat{a}_1^\dag+\hat{a}_1)\,(\hat{a}_2^\dag+\hat{a}_2).
\end{align}
Here we have defined the coupling constants $g_\text{C}:=\tilde{C}_\textbf{c}\,p_{1\text{ZPF}}\,p_{2\text{ZPF}}$ and $g_\text{L}:=1/\tilde{L}_\textbf{c}\,\phi_{1\text{ZPF}}\,\phi_{2\text{ZPF}}$.

Notice that, as was the purpose of this derivation, we have shown that the interaction part of the Hamiltonian \eqref{hamiltonioan:CQED:appendix} matches exactly the terms presented in \eqref{circuit:couplings}.

\end{document}